\documentclass[useAMS]{mn2e}
\usepackage{psfig}

\newif\ifAMStwofonts

\def\be{\begin{equation}}
\def\ee{\end{equation}}

\def\etal{{\it et al.~}}

\def\gtsima{$\; \buildrel > \over \sim \;$}
\def\ltsima{$\; \buildrel < \over \sim \;$}
\def\prosima{$\; \buildrel \propto \over \sim \;$}
\def\gsim{\lower.5ex\hbox{\gtsima}}
\def\lsim{\lower.5ex\hbox{\ltsima}}
\def\simgt{\lower.5ex\hbox{\gtsima}}
\def\simlt{\lower.5ex\hbox{\ltsima}}
\def\simpr{\lower.5ex\hbox{\prosima}}

\def\etal{{\frenchspacing\it et al. }}
\def\ie{{\frenchspacing\it i.e. }}
\def\eg{{\frenchspacing\it e.g. }}

\def\be{\begin{eqnarray}}
\def\ee{\end{eqnarray}}

\def\CR{{\tt CRASH}~}
\def\RT{radiative transfer~}

\title[CRASH: a  Radiative Transfer Scheme]{CRASH: a  Radiative Transfer Scheme}

\author[A. Maselli, A.Ferrara and B.Ciardi]{
A. Maselli$^{1}$,
A. Ferrara$^{2}$
\&
B. Ciardi$^{3}$\\
$^1$ Dipartimento di Astronomia, Universit\'a di Firenze,  Largo
Enrico Fermi 5, 50125 Firenze, Italy\\
$^2$ International School for Advanced Studies, SISSA, via Beirut 2-4,
34013 Trieste, Italy\\
$^3$ Max-Planck-Institut f\"ur Astrophysik, Karl-Schwarzschild-Strasse 1, 
85748 Garching, Germany
}

\date{April 2003}

\pagerange{\pageref{firstpage}--\pageref{lastpage}}
\pubyear{2003}
\begin{document}

\maketitle
\label{firstpage}

\begin{abstract}                                                                    
We present a largely improved version of {\tt CRASH},
a 3-D radiative transfer code that treats the effects 
of ionizing radiation propagating through a given       
inhomogeneous H/He cosmological density field, on the physical 
conditions of the gas.  
The code, based on a Monte Carlo technique, self-consistently calculates the 
time evolution of gas temperature and ionization fractions
due to an arbitrary number of point/extended sources and/or 
diffuse background radiation with given spectra. 
In addition, the effects of diffuse ionizing radiation following
recombinations of ionized atoms have been included.  
After a complete description of the numerical scheme,
to demonstrate the performances, accuracy, convergency and robustness of the code 
we present four different test cases designed to investigate 
specific aspects of radiative transfer: (i) pure hydrogen isothermal Str\"omgren sphere 
; (ii) realistic Str\"omgren spheres; (iii) multiple
overlapping point sources, and (iv) shadowing of background radiation 
by an intervening optically thick layer.
When possible, detailed quantitative comparison of the results against 
either analytical solutions or 1-D standard photoionization codes
has been made showing a good level of agreement. 
For more complicated tests the code yields physically plausible results,
which could be eventually checked only by comparison with other similar codes.
Finally, we briefly discuss future possible developments and 
cosmological applications of the code.
\end{abstract}

\begin{keywords}
cosmology: radiative transfer - intergalactic medium -
 Ly$\alpha$ forest - reionization
\end{keywords}

\section{Introduction}
Stimulated by the growing number of observations
of the high-redshift universe (\eg Fan \etal 2001, 2003; Hu \etal 2002),
an increasing attention
has been dedicated to the theoretical modeling of the reionization
process, adopting both
semi-analytic and numerical approaches (\eg Gnedin \& Ostriker 1997;
Haiman \& Loeb 1997; Valageas \& Silk 1999; Ciardi \etal 2000, CFGJ;
Miralda-Escud\'e, Haehnelt \& Rees 2000; Chiu \& Ostriker 2000;
Gnedin 2000; Benson \etal 2001; Razoumov \etal 2002;
Ciardi, Stoehr \& White 2003, CSW).
An important refinement
has been introduced by the proper treatment of a number of feedback effects
ranging from the mechanical energy injection to the H$_2$ photodissociating
radiation produced by massive stars (CFGJ).
Only recently, though,  the reionization process has been studied in
a cosmological context including structure evolution, a realistic
galaxy population and a sufficiently accurate treatment of the Radiative
Transfer (RT) of ionizing photons through the InterGalactic Medium (IGM),
in a simulation volume large enough to have properties ``representative''
of the all universe (CSW).
The next challenge for this kind of study is to develop accurate and fast
radiative transfer schemes that can be then implemented in cosmological simulations.

Although the reionization process is one of the most demanding and natural applications
of the radiative transfer theory,
numerous physical problems require a detailed understanding of the
propagation of photons into different environments, ranging
from intergalactic and interstellar medium to stellar or planetary
atmospheres. The full solution of the seven dimension RT
equation (three spatial coordinates, two angles, frequency and time)
is still well beyond our computational capabilities
and, although in some specific cases it is possible to reduce its
dimensionality, for more general problems no spatial symmetry
can be invoked. Thus, an increasing effort has been devoted to the
development of radiative transfer codes based on a variety of approaches
and approximations (\eg Umemura, Nakamoto \& Susa
1999; Razoumov \& Scott 1999; Abel, Norman \& Madau 1999; Gnedin 2000;
Ciardi et al. 2001, CFMR; Gnedin \& Abel 2001; Cen 2002).
In the following, we give a list of the available cosmological radiative
transfer codes, together with their characteristics and applications.

\begin{itemize}
\item {\it Gnedin \& Ostriker (1997)}: time-dependent code based
on the local optical depth approximation. It solves the evolution
of H, He, H$_2$ and temperature. It includes the RT of radiation from 
multiple point sources, a background flux and the diffuse component
due to recombination. The code is fast, independent
on the number of sources, but the approximation breaks down for
optical depths of the order of one and cannot account for shadowing
effects.  It has been applied to the study of the reionization
process and has been coupled with hydrodynamical simulations in
the implementation of Gnedin (2000).
\item {\it Umemura, Nakamoto \& Susa (1999)}: time-independent
code based on the method of the short characteristics. It calculates
the equilibrium configuration of an isothermal H/He gas.
It includes the RT of a background flux and the diffuse component
due to recombination. As a backdraw the code can calculate 
only equilibrium configurations. It has been applied
to the study of cosmological reionization.
\item {\it Abel, Norman \& Madau (1999)}: time-dependent ray-tracing
code. It solves the evolution of hydrogen and
treats RT of the radiation emitted by a point source, together with the
diffuse component. Although the ray-tracing approach is very
accurate, its primary limitation is the high computational cost
required to follow a large number of sources and to assure the
correct coverage of the simulation volume. This code has been
applied to the study of the evolution of the ionized region surrounding
a mini-quasar in a cosmological density field.
\item {\it Razoumov \& Scott (1999):} time-dependent ray-tracing code,
based on the solution of the 5D (three spatial coordinates and two angles)
advection equation. It includes the H, He and H$_2$ chemistry and the
temperature evolution. It treats the RT of radiation from
multiple point sources, a background flux and the diffuse component. 
The code suffers from the same problem as the previous one.  
It has been applied to the study of the reionization process. 
\item {\it Ciardi \etal (2001):} \CR 1.0 time-dependent code, based on a Monte
Carlo approach. It includes the evolution of an isothermal H gas and
it treats the RT of the radiation emitted by point sources, together with the
diffuse component. The code has been applied to the study of the evolution of
the ionized region surrounding point sources embedded in a
cosmological density field.
\item {\it Gnedin \& Abel (2001):} time-dependent code, based on
the explicit moment solver OTVET (Optically Thin Variable Eddington Tensor
formalism). It includes the H and He chemistry and the temperature evolution.
It treats the radiation produced by point sources, a background
flux and the diffuse component. Its main problem
is related to the correct evaluation of the Eddington tensors. In fact,
while the assumption of optically thin Eddington tensor is exact in the
limit of a single point source, it may break down in complex situations
with very inhomogeneous source functions and opacity fields. It has
been coupled with a hydrodynamical simulation to study cosmological 
reionization.
\item {\it Sokasian, Abel \& Hernquist (2001):} time-dependent
ray-tracing code. It includes the H and He chemistry and the temperature evolution.
It treats the radiation produced by point sources, a background
field and the diffuse component. As a first application, it has been 
used to the compute HeII reionization by quasars.
\item {\it Cen (2002):} time-dependent ray-tracing code, based on the
fast Fourier transform method for hydrogen ionization.
It treats the radiation produced by point sources, a background
field and the diffuse component. Unlike conventional
ray-tracing schemes where angular discretization is performed on the
source, here it is done at the receiving site, guaranteeing good space coverage
regardless of the finesse of the angular discretization. Moreover,
its speed is independent on the number of sources. The code has been applied to
test cases of single and multiple sources embedded in a constant or
inhomogeneous density field.
\item {\it Razoumov \etal (2002):} time-dependent ray-tracing code.
It includes the H, He and H$_2$ chemistry and the temperature evolution.
It treats the radiation produced by point sources, a background
field and the diffuse component. To alleviate the speed problem related 
to the ray-tracing approach, the
authors have implemented the method developed by Abel \& Wandelt (2002),
introducing trees of rays segments which recursively split
into sub-segment as one moves away from the source. The code has been
applied to the study of the reionization process.
\end{itemize} 

Monte Carlo (MC) methods, on which we base our approach, have been
widely used in several (astro-)physical
areas to tackle RT problems (for a reference book see
Cashwell \& Everett 1959) and they have been shown to result in fast and accurate
schemes. Here we build up on previous work (Bianchi,
Ferrara \& Giovanardi
1996; Ferrara \etal 1996; Ferrara \etal 1999; Bianchi \etal 2000)
to develop a new version of the code {\tt CRASH} ({\it Cosmological Radiative
transfer Scheme for Hydrodynamics}), which improves in many ways the earlier
version presented in CFMR.
The main aim of this paper is to give a full description of the code
\CR in its present form, perform a number of stringent tests that demonstrate its
ability to treat RT problems arising in a wide range of astrophysical
context, and evaluate its performances. In forthcoming work, we will
apply the code to specific cosmological problems.

The plan of the paper is as follows.
In Sec. 2 we describe the details of the numerical implementation concerning
the discretization of the radiation field and its (time-dependent) interaction
with the surrounding matter; we pay special attention to issues as the
dependence of spatial and time resolution on the numerical parameters.
In Sec. 3, we present results from a set of specifically designed test-runs:
Str\"omgren spheres, two point sources and an optically
thick slab case illuminated by a background field.
Finally, in Sec. 4, we draw our conclusions and discuss some
of the possible future applications and developments of \CR.

\section{Description of the Code}

\CR, 
is a 3-D numerical code primarily   developed to study time-dependent \RT 
problems in cosmology; however, the versatility of its scheme allows
easy extensions to additional astrophysical applications.
Such a scheme is largely based on MC techniques, which 
are used to sample appropriate Probability Distribution Functions (PDFs), 
as briefly explained in Appendix A.
The current version of the code (\CR 2.0) is a development of a previous
implementation of the same scheme, discussed in CFMR. 
The main differences with that first attempt are [i] the inclusion
of helium in all its ionization states, [ii] the self-consistent 
calculation of gas temperature, and [iii] an efficient algorithm to 
treat multiple radiation sources and/or a diffuse background field.
In addition, more refined photon propagation schemes (see below) 
have been included.  
As today, \CR works as a stand-alone routine that can be applied to 
arbitrary and pre-computed hydrodynamical density fields.  
A future development will consist of a full coupling with 
available cosmological hydrodynamical codes, so that 
the interaction of \RT and gas dynamics can be treated self-consistently. 
We now turn to the description of the implementation of the physical
processes and the numerical algorithms used by the code.  

\subsection{Initial Conditions}

As a first step, the initial conditions for the physical quantities
(gas density, temperature, H and He ionization fractions) 
must be assigned on the $N_c^3$ nodes of the adopted 3-D cartesian grid.  
The density field, the temperature and the ionization fractions 
can be either the output of a hydrodynamical 
simulation or arbitrarily constructed. While the density field remains 
constant during the simulation, \ie no back-reaction of gas dynamics to the 
effects of the \RT is considered, the temperature and ionization fractions
are instead updated with time. Both open and periodic
boundary conditions can be adopted. In the latter case, the ionizing 
radiation that exits with a given direction from a cell $(x_b,y_b,z_b)$ 
on the boundary surface of the simulation box, is re-emitted with the same 
direction from the cell $(-x_b,-y_b,-z_b)$; 
the origin of the cartesian coordinate system is 
located at the box center. 

Finally, the properties of the radiation field must be specified. 
The quantities required are the number, location and emission
properties (\eg the intensity of the emitted radiation and its
spectral energy distribution) of point sources in the simulated volume, 
and the intensity and spectral energy distribution of the background radiation, 
if present. As \CR has been specifically designed to follow the propagation of
ionizing photons, the above spectral parameters are required only for energies
$\ge 1$~Ryd. The required properties of the radiation fields are described
in detail in the next Section.

\subsection{Ionizing Radiation Field}

The MC approach to \RT requires 
that the radiation field is discretized into photon packets;
the properties of the radiation field and the relevant radiation-matter 
interaction processes are then statistically treated by randomly sampling 
appropriate PDFs. In
this approach the ionizing 
radiation emitted by either a specified number of point sources located 
arbitrarily in the box, or a background radiation, is easily
reproduced. Also, 
diffuse radiation from recombinations in the ionized gas can be 
treated in a similar manner without further assumptions.

\subsubsection{Point Sources}
Let us first consider the case of a single point source with a time-dependent
bolometric luminosity $L_s(t)$. 
If $t_s$ is the (physical) simulation time, the total energy emitted by
the source is: 
\be
\label{eq01} 
E_s= \int_0^{t_s} L_s(t)dt.
\ee
Such energy is distributed into $N_p$ photon packets, emitted 
at the source location
at regularly spaced time intervals $dt=t_s/N_p$. Thus, 
the time resolution of a given run is fixed by $N_p$.
To each emitted photon packet a frequency, $\nu$, and a propagation direction, 
$(\theta,\phi)$, are assigned by sampling the source spectrum, $S_{\nu}$, 
and the angular PDF, using the MC method described in Appendix A. 
This procedure allows to treat
any given spectrum and possible emission anisotropies.  
The $j$-th monochromatic packet of frequency $\nu$ is then emitted at time $t_j=jdt$ 
($j=1$,.....,$N_p$), and it contains $N_{\gamma,j}= \Delta E_j/h\nu$ photons, 
where $\Delta E_j$ is the energy of the $j$-th packet given by: 
\be
\label{eq02} 
\Delta E_j= \int_{t_{j-1}}^{t_{j}} L_s(t)dt.
\ee
Once the quantities $\nu$, ($\theta,\phi$) and $N_{\gamma,j}$ are assigned to an
emitted photon packet, we follow its propagation from the source location
into the given density field as described in Sec.~2.3. 
 
\subsubsection{Background Radiation}
 
The presence of an external ionizing background radiation 
is treated with an algorithm analogous to the one used for point 
sources, with the only difference that the emitting location
is one of the $6\times N_c^2$ cells on the faces of the simulation box,
rather than the cell corresponding to the source location.
If the background radiation is uniform and isotropic, 
an integer number is randomly extracted in $[1,2,3,4,5,6]$ to select a face 
of the box and two 
random integer numbers in  [$1, N_c$] to select the cell position on the 
chosen face. This procedure can be easily generalized to account for
possible asymmetries in the radiation field.  
 
\subsubsection{Diffuse Radiation} 
 
Diffuse ionizing radiation in the IGM can 
be produced by recombinations and 
by free-free (f-f) transition processes. 
Nevertheless the f-f emissivity at frequencies higher than 1 Ryd 
is negligible with respect to the emissivity due to recombinations 
and we can safely exclude this process. 
The implementation includes as 
sources of diffuse radiation the recombination processes 
occuring in a H/He gas:
\begin{itemize}
\item{}H$^+$ + e$^-$ $\longrightarrow$ H$^0$ + $h\nu$,
\item{}He$^+$ + e$^-$ $\longrightarrow$ He$^0$ + $h\nu$,
\item{}He$^{++}$ + e$^-$ $\longrightarrow$ He$^+$ + $h\nu$.
\end{itemize}
The MC scheme allows a straightforward and self-consistent 
treatment of the diffuse radiation produced by H/He recombinations 
in the ionized gas.    
In analogy with the direct one, the diffuse radiation field is 
discretized into photon packets whose propagation is followed 
using the scheme discussed in the following Section. 
For this approach, it is necessary to assign the angular and 
frequency distribution of the emitted diffuse radiation 
as well as its intesity. 

As the emission process for diffuse radiation is isotropic, 
the packet propagation direction is randomly selected. 
The spectral distribution of the diffuse radiation depends 
on the emitting species. If LTE is assumed, 
the emissivity $\eta(\nu)$ associated to an arbitrary 
recombining atom, results in the following 
frequency dependence (Mihalas 1978, Osterbrok 1989):
\be
\label{eq05} 
\eta_{\cal H}(\nu) \propto \sigma_{\cal H}(\nu) \nu^3 
e^{-(h\nu-h\nu_{th,{\cal H}})/k_B T}  
\;\;\;\;\;\;\;\;\;\;\;\; \nu \ge \nu_{th,{\cal H}},
\ee
where $\sigma_{\cal H}(\nu)$ and $h\nu_{th,{\cal H}}$ are the 
photoionization cross-section and the ionization potential of 
the recombined atom ${\cal H}$; $k_B$ is the Boltzmann constant and
$T$ is the kinetic temperature of the recombining electron. 
In the calculation we include only the effect of the ionizing 
diffuse radiation.

The intensity of the diffuse radiation is evaluated as follows. 
For each of the three species we allocate
a 3-D array, $N_{rec}(x,y,z)$, composed of $N_c^3$ elements, \ie one for
each cell of the 3-D cartesian grid. Each element of $N_{rec}(x,y,z)$ 
keeps track of the number of recombination events occurred in the 
corresponding cell. 
For simplicity, we now consider a fixed cell and the recombination process of
species $I$ ($I \in \{{\rm H}^+, {\rm He}^{+}, {\rm He}^{++}\}$)
to describe how $N_{rec}(x,y,z)$ is computed. 
If a packet crosses the fixed cell at a time $t_c$, the number of 
recombinations, $\Delta N_{rec}$, occurred since the last time, 
$t_{c-1}$,  a packet crossed the same cell is approximately given by: 
\be
\label{eq03} 
\Delta N_{rec} \simeq \alpha({\tiny T_{c-1}}) 
n_{e,c-1} n_{I,c-1} \Delta t (\Delta x)^3,
\ee
where $T_{c-1}, n_{e,c-1}$ and $n_{I,c-1}$ are the temperature, 
electron number density and number density of species $I$
evaluated at time $t_{c-1}$; $\alpha({\tiny T_{c-1}})$ is the total 
recombination coefficient of species $I$, $\Delta t=t_c-t_{c-1}$ is 
an integer multiple of $dt$, and $\Delta x$ the linear dimension of 
a cell. We then update the number of recombinations in the cell, 
$N_{rec,c}=N_{rec,c-1}+\Delta N_{rec}$. If $N_{rec,c}$
satisfies the following condition:  
\be
\label{eq04} 
N_{rec,c} > f_r  N_a,
\ee
a packet containing $N_\gamma=N_{rec,c}$ photons is emitted from the cell and
$N_{rec,c}$ is set to zero; we assume that all $N_{rec,c}$ recombinations 
occur to the same atomic level, so that the emitted packet is monochromatic. 
In the above equation, $f_r \in [0,1]$ is an adjustable numerical parameter 
and $N_a$ is the total number of recombining atoms 
in the cell, \ie $n_{\rm H}(\Delta x)^3$ or $n_{\rm He}(\Delta x)^3$, 
where $n_{\rm H}=n_{\rm H^0}+n_{\rm H^+}$ and $n_{\rm He}=n_{\rm He^0}+
n_{\rm He^+}+n_{\rm He^{++}}$. 
In principle low values of $f_r$ make the simulated recombination 
process smoother at the expense of a higher computational cost; 
experimentally we have found that $f_r=0.1$ represents a good 
compromise between accuracy and computational cost.  
 
At each recombination event, {\it e.g.} each time the condition \ref{eq04} 
is satisfied, we stochastically derive the probability for recombinations at ground level
to occur from the ratio  $\alpha_1(T)/\alpha(T)$, with 
$\alpha_1$ being the recombination coefficient to the ground level 
(the expressions for all the rates adopted in \CR are given in Appendix B). 
If recombination occurs to the first level we sample 
the spectrum in eq.~\ref{eq05} (appropriately normalized) for all the 
three reemission processes. Recombinations to levels 
other than the first of H$^0$ are neglected as they 
produce no ionizing photons. 
De-excitations following recombinations to higher levels  
of He$^0$ or He$^+$ produce instead ionizing photons for H$^0$ or He$^0$, 
respectively. The de-excitation can occur through different channels; 
however for this cases the frequency assigned to the emitted packet 
is  the one corresponding to the relevant transition from the second to 
the ground level: for He$^0$ we assume the mean value between 
$h\nu=19.8$~eV and $h\nu=21.2$~eV (2s$\rightarrow$1s 
and 2p$\rightarrow$1s); for He$^+$ $h\nu=40.7$eV (2p$\rightarrow$1s). 
This approximation has negligible effects on the results of a typical 
cosmological simulation and it can be easily modified if a more 
detailed treatment of chemistry is necessary.

\subsubsection{Multiple Sources}
 
As already pointed out, \CR  allows to run simulations with more than one source. 
Let us consider the case of $N_s$ sources (one of which could be a background radiation
field) 
emitting simultaneously ionizing radiation inside the box. 
Each of the $N_s$ sources is treated as described in the previous Sections,   
assuming the same value of $N_p$ for all of them. In this way it is possible to 
reproduce the time evolution of the total ionizing radiation field 
in the computational volume, by emitting the $j$-th packet ($j \in \{1,..,N_p\}$) 
from all the $N_s$ sources at the same time-step 
$t_j=j dt$, looping on the sources at each $j$ value.

This simple method has been chosen as it is computationally very cheap. 
We have already noted that the time resolution and  
the accuracy (and the convergency, see later on)
of the simulation depends on the total number of packets emitted
by a single source, $N_p$. 
In general, to achieve the same accuracy level when multiple sources are
present, one would need to multiply such number by the number of sources
in the box, thus becoming $N_p N_s$, with a significant increase of the 
computational time. 
However, in practice this 
is not necessary because, unless either the ionized regions created by 
each source are completely separated and/or the ambient gas is very 
optically thick, almost inevitably a typical cell in the computational 
domain will ``see'' a large number of sources, hence the effective number
of photon packets required is $\ll N_p N_s$. 
\subsection{Photon Packets Propagation}
The MC approach to the RT problem is particularly convenient. 
In fact, as the radiation is modeled in its particle-like nature, it is not necessary 
to solve the high-dimensional cosmological RT                 equation for the 
specific intensity of the radiation, which usually requires several 
analytical and numerical approximations (Abel et al. 1999; Razoumov \& Scott 1999; 
Nakamoto et al. 2001; Razoumov et al. 2002).

To describe the propagation direction of the emitted packets we
adopt spherical coordinates, $(r, \theta ,\phi)$, 
with origin at the emission cell, $(x_e, y_e, z_e)$. Thus, a packet will propagate along
the direction identified by:
\be
\label{eq06}
\left[
\matrix{
x & = &x_e+ r \sin \theta \cos \phi  \cr
y & = &y_e + r \sin \theta \sin \phi \cr
z & = &z_e + r \cos \theta \cr
}
\right]
\ee
Here, ($\theta,\phi)$ are determined via the MC method applied to
the appropriate emission angular PDF of the source (see Sec.~2.2.1).      
The cells crossed by the packet can be ordered by the index $l=0,...,l_{max}$
(where $l_{max}$ is the maximum number of cells that are crossed by the packet), 
which is related to the cells cartesian
coordinates by the following expression:
\\
\begin{eqnarray}
\label{eq07}
x (l)& = &x_e+ int (l\; a_0) \nonumber \\
y (l)& = &y_e + int (l \;b_0) \\
z (l)& = &z_e + int (l \;c_0). \nonumber
\end{eqnarray}
Here $a_0$, $b_0$ and $c_0$ are the director cosines of the propagation direction
with respect to the cartesian grid: 
$a_0= \sin \theta \cos \phi$, $b_0= \sin \theta \sin \phi$, $c_0= \cos \theta$.

Given the above definitions, let us now focus on the propagation of the generic
monochromatic packet initially composed of $N_\gamma$ photons of frequency $\nu$.
As already mentioned, photoelectric absorption of continuum ionizing 
photons is the only opacity contribution included.
Thus, the absorption probability for a single photon travelling 
through an optical depth
$\tau$ is given by:
\be
\label{eq08} 
P(\tau)= 1 - e^{-\tau}.
\ee
As the packet propagates through the above cell sequence, 
it will deposit in the $l$-th cell a fraction of its initial 
photon content $\propto P(\tau^l)$, where $\tau^l$ is the  
total optical depth crossed from the emission site to the cell itself.  
For each cell crossed, we compute the contribution, $\Delta^l \tau$, 
of the actual cell to the total optical depth, as follows:
\be
\label{eq09} 
\Delta^l \tau=\Delta^l \tau_{H^0}+\Delta^l \tau_{He^0}+\Delta^l \tau_{He^+}
= [\sigma_{H^0}(\nu) n_{H^0}^l + \nonumber \\
\sigma_{He^0}(\nu)  n^l_{He^0} +
     \sigma_{He^+}(\nu) n_{He^+}^l] f(l)\Delta x, 
\ee
where $\sigma_A$ is the photoionization cross-section for 
absorber $A \in \{ H^0,$ $He^0$, $He^+\}$, $n_A^l$ its numerical 
density in $l$-th cell and $f(l)\Delta x$ the path length 
through the $l$-th cell of linear size $\Delta x$. 
Depending on the trajectory of the packets in the cell, $f(l)$ will 
have a value in the range $[0, \sqrt{3}]$.
To limit the computational cost, we do not include in the code a 
ray-casting routine to determine the path length in each cell. 
We assume, instead, the fixed value $f(l)=0.56$,
corresponding to the median value of the PDF
for the lengths of randomly oriented paths entering from the faces of 
a cubic box of unit size (see CFMR). \\
The number of photons absorbed in the $l$-th cell, $N_A^l$, is determined as
follows.
When a packet reaches the $l$-th cell, its photon content is:                  
\be
\label{eq11} 
N_\gamma^l= (N_\gamma^{l-1}-N_A^{l-1})= N_\gamma^{l-1} e^{-\Delta^{l-1} \tau} 
\le N_\gamma;
\ee
it will deposit in the same cell a number of photons equal to:
\be
\label{eq12} 
N_A^l=N_\gamma^{l}(1-e^{-\Delta^l \tau}).
\ee
In some cases $N_A^l$ can exceed the total number of atoms
that can be ionized in the $l$-th cell, $N_{ion}^l$; when this occurs 
we set $N_A^l=N_{ion}^l$ and retain the excess photons in the traveling
packet.   
Propagation of each packet is followed until it exits from the box 
(open boundary condition case)   or until 
extinction occurs when $N_\gamma^l < 10^{-p} N_\gamma$. This procedure
guarantees energy conservation to an equivalent accuracy of $10^{-p}$. 
We typically adopt $p \in [4,9]$ depending on the required accuracy.
\subsection{Updating Physical Quantities}
In this Section we describe the scheme adopted to calculate the time evolution 
of the ionization fractions ($x_{H^+}=n_{H^+}/
n_H$, $x_{He^+}=n_{He^+}/n_{He}$, $x_{He^{++}}=n_{He^{++}}/n_{He}$) and of 
the gas temperature, $T$. 
The system of coupled equations that describes
the above evolution is the following:
\be
\label{eq13} 
n_{\rm H} \frac{dx_{H^+}}{dt}&=&\gamma_{H^0}(T) n_{H^0} n_e - 
\alpha_{H^+}(T)n_{H^+}n_e  \nonumber  \\
&&+ \;\Gamma_{H^0}n_{H^0}={\cal I}_{H^+}, \nonumber  \\
n_{\rm He} \frac{dx_{He^+}}{dt}&=&\gamma_{He^0}(T) n_{He^0}n_e   - 
\gamma_{He^+}(T) n_{He^+}n_e  - \nonumber  \\ 
&&\alpha_{He^+}(T)n_{He^+}n_e +\alpha_{He^{++}}(T)n_{He^{++} n_e }\nonumber  \\
&&+\; \Gamma_{He^0}n_{He^0}={\cal I}_{He^+},  \\
n_{\rm He} \frac{dx_{He^{++}}}{dt}&=&\gamma_{He^+}(T) n_{He^+} n_e- 
\alpha_{He^{++}}(T)n_{He^{++}} n_e\nonumber \\
 && +\; \Gamma_{He^+}n_{He^+}={\cal I}_{He^{++}}, \nonumber \\
\frac{dT}{dt}&=&\frac{2}{3k_Bn}\left[k_BT\frac{dn}{dt}+ {\cal H} (T, x_I ) 
- \Lambda(T, x_I)\right]. \nonumber 
\ee
The last equation is the energy conservation equation, where
$n=n_{\rm H}+n_{\rm He}+n_e$ is the number of free particles per unit 
volume in the gas; ${\cal H}$ and $\Lambda$ are the heating 
and cooling functions which account for the energy gained 
and lost from the gas in the unit volume per unit time, 
respectively (see Appendix B for the detailed $\Lambda$ 
expression adopted). 
In the three equations for the ionization fractions we indicate with 
$\alpha_I$ ($\gamma_A$) the recombination (collisional ionization)
coefficient and $I \in \{$H$^+$, He$^+$, He$^{++} \}$ ($A \in \{$H$^0$, He$^0$, He$^{+} \}$);
$\Gamma_A$ is the time-dependent photo-ionization rate.

The numerical approach requires to discretize the differential equations 
as follows:   
\be
\label{eq15}
x_{H^+}(t+\Delta t) &=& x_{H^+}(t) +{\cal I}_{H^+}(t)\Delta t/n_{\rm H} , \nonumber\\
x_{He^+}(t+\Delta t) &=&  x_{He^+}(t) +{\cal I}_{He^+}(t)\Delta t/n_{\rm He} ,\nonumber\\
x_{He^{++}}(t+\Delta t)&=&x_{He^{++}}(t) +{\cal I}_{He^{++}}(t)\Delta t/n_{\rm He} ,\\
T(t+\Delta t)&=& T(t)+\frac{2}{3k_Bn}\big\{k_BT \Delta n + \nonumber\\
&&\Delta t [{\cal H} (T, x_I )
- \Lambda(T, x_I)]\big\},  \nonumber
\ee
where $\Delta n=n(t+ \Delta t)-n(t)$. 
The physical quantities in a cell are updated by solving the above system each time
the cell is crossed by a packet. Thus, the integration time step $\Delta t$ 
(\ie the time interval 
between two subsequent passages of a packet in a given cell) is not constant, due to the 
statistical description of the emission and propagation processes. 
The integration time step is calculated as follows. 
We introduce a 3-D array whose $N_c^3$ elements correspond to the index of the last
packet that has crossed a cell (Sec.~2.2.1). Thus,  
when the $j$-th packet crosses the cell with coordinates $(x,y,z)$ we update the
physical quantities in the cell, using an integration time step equal to:      
\be
\label{eq16} 
\Delta t = \big[j-j'(x,y,z)\big] dt,
\ee
where the array element $j'(x,y,z)$ gives the index of the packet 
last transited through $(x,y,z)$, and we update the array element to the actual 
index $j$.

As radiation field is discretized in photon packets, it is not straightforward 
to recover continuous quantities that directly depend on its intensity,
as photo-ionization and photo-heating rates. 
Their effects are evaluated by means of 
discretized contributions as a function of 
the number of photons deposited in the cell, 
$N_A^l=N^l_{A,H^0}+N^l_{A,He^0}+N^l_{A,He^+}$; these are distributed 
among the three absorbing species
proportionally to their absorption probability 
$(1-e^{-\Delta^l \tau_A})$. 
The corresponding ionization fractions and temperature increases are given by:

\be
\label{eq17}
\Delta x_{H^+} &= & \frac{n_{\rm H^0}}{n_{\rm H}} \Gamma_{H^0}\Delta t\equiv
\frac{N_{A,H^0}^l}{N^l_H}\nonumber \\
& = &\frac{N_A^l}{n^l_H \Delta^3 x}
\left( \frac{1-e^{-\Delta \tau^l_{H^0}}}{1-e^{-\Delta \tau^l}}\right) , \nonumber \\
\Delta x_{He^+} &= & \frac{n_{\rm He^0}}{n_{\rm He}} \Gamma_{He^0}\Delta t \equiv
\frac{N_{A,He^0}^l}{N^l_{He}}\nonumber \\
& = & \frac{N_A^l}{n^l_{He} \Delta^3 x}
\left( \frac{1-e^{-\Delta \tau^l_{He^0}}}{1-e^{-\Delta \tau^l}}\right) ,  \\
\Delta x_{He^{++}} &= & \frac{n_{\rm He^+}}{n_{\rm He}} \Gamma_{He^+}\Delta t \equiv
\frac{N_{A,He^+}^l}{N^l_{He}}\nonumber \\
& = & \frac{N_A^l}{n^l_{He} \Delta^3 x}
\left( \frac{1-e^{-\Delta \tau^l_{He^+}}}{1-e^{-\Delta \tau^l}}\right) ,\nonumber \\
\Delta T &= &\frac{2}{3k_Bn}\{k_B\Delta n +[N_{A,H^0}^l (h\nu-h\nu_{th,H^0})+ \nonumber \\
&&N_{A,He^0}^l (h\nu-h\nu_{th,He^0})+ \nonumber\\
&&N_{A,He^+}^l (h\nu-h\nu_{th,He^+}) ] \}. \nonumber
\ee
Recombinations, collisional ionizations and cooling are instead treated 
as continuous processes.
This requires $\Delta t$ to be much smaller 
than the characteristic timescales of these processes for all species 
$I$ and $A$, \ie $\Delta t \ll t_{min}=min\{t_{rec,I},t_{coll,A},t_{cool}\}$. 
If this condition is not fulfilled, the integration is split into $n_s$ steps, with 
$n_s=int\left[\Delta t / (f_s t_{min})\right]$; $f_s$ is 
a fudge factor, usually taken equal to $f_s=50-100$, in order to minimize 
the discretization errors.  
The expressions used for recombination, collisional ionization and cooling rates    
are given explicitly in Appendix~B. 
\subsection{Numerical Resolution}
The input parameters that determine the spatial/time resolution 
and the computational cost of a simulation are the following:
\begin{itemize}
\item $N_p$: number of photon packets emitted per source. 
\item $N_c^3$: number of grid cells. 
\item $N_s$: number of sources in the simulation.
\end{itemize}   
Although physical processes in our scheme are treated statistically,  
it is nevertheless possible to derive semi-empirical dependences 
among these parameters and the performances of \CR in terms of 
resolution and speed. 
Once the initial conditions are fixed, $N_p$ determines 
the mean number of photons contained in a single packet, $N_\gamma$ (actually the 
photon number content in each packet depends on its energy and frequency), 
and the time interval between two subsequent packet emissions, $dt$. 
The number of grid cells, $N_c^3$, determines the spatial resolution and 
the mean number of atoms in each cell, $N_a$.

Let us consider the general case of a simulation with $N_s$ sources and open
boundary conditions. The total number of mitted packets is 
$N_p^{tot}=N_s N_p$. To a first order of magnitude a packet will pierce a mean number of cells equal 
to $f_d N_c$, where $f_d$ is a parameter ranging in $[{N_c}^{-1},1]$: 
$f_d \approx 1$ for an optically thin medium and $f_d \approx {N_c}^{-1}$ for a 
simulation box with optically thick cells.
During a simulation a cell will be crossed and updated 
an average number of times equal to 
\be
\label{eq18} 
N_{cr}=  \frac{N_p^{tot} f_d N_c}{N_c^3}=f_d \; \frac{N_s N_p}{N_c^2};
\ee
the accuracy of a given simulation is mainly determined  by the magnitude of
this parameter. 
In order to correctly reproduce the physics of photoionization, in fact, 
the value of $N_{cr}$ must be large enough to appropriately sample
the frequency  and the time distribution of the ionizing 
radiation at each cell location. This implies that it is necessary
to impose a lower limit on $N_{cr}$. We find that 
most common (\eg power-law, black-body) spectra can be well 
MC sampled with roughly $10^2-10^4$ extractions ({\it i.e.} packets crossing each cell), 
depending on the shape of the spectra and
on the required accuracy. Hence we require that $N_{cr}$ exceeds the previous 
estimate, to achieve a good spectral sampling in each cell.
A further condition has to be imposed on $N_{cr}$, which 
determines also the mean value of the integration time-step, 
$\langle\Delta t\rangle= {t_s / N_{cr}}$ (see above).
As already pointed out, $\Delta t$ must be much smaller than the 
characteristic time scales of the physical processes treated as 
continuous. This imposes an upper limit to $\Delta t$, 
or, in turn, a lower limit to $N_{cr}$ set by: 
\be
\langle\Delta t\rangle=\frac{t_s}{N_{cr}} \ll t_{min} \Rightarrow  
N_{cr}=f_d\frac{N_s N_p}{N_c^2} \gg \frac{t_s}{t_{min}}.
\label{eq19}
\ee 
We initialize $t_{min}=min\{t_{rec,I}, t_{coll,A}, t_{cool}\}$
before the run, with the characteristic times calculated using the mean 
values of the initial temperature and ionization fractions fields; 
although this quantities differ from cell to cell and will inevitably change
due to the evolution of the physical properties of the gas during the simulation. 
However, if the condition eq.~\ref{eq19} is not fulfilled at some computational 
stage, we resort again to splitting as described in Sec.~2.4. 
Before running a simulation, the numerical parameters have to be chosen 
such to satisfy the above conditions; then the accuracy of the 
calculation increases with $N_{cr}$. 

Two other combination of the input parameters influence the accuracy of the calculation. The first one 
is the ratio $N_\gamma/N_a$ in a single cell which determines the 
mean number of packets 
necessary to completely ionize a single cell; the lower its value, the better the adopted 
scheme is able to reproduce the continuity of the photoionization (photoheating) 
process.
 
The second combination 
is the one giving the ratio between the photoionization rate and 
that of continuous processes (\eg recombinations) averaged in the simulation box. 
Taking into account the geometrical dilution of the ionizing radiation 
emitted by point sources, a rough estimate for such quantity  is 
$<\dot N_\gamma> (L_{box})^{-2} t_{min}$, where $<\dot N_\gamma>$ is the mean ionizing 
photon rate. 
As discussed in the previous Section, we adopt a different method to  describe the 
effects of these two classes of physical processes. The scheme adopted to account for
photoionization is very precise, as it guarantees energy 
conservation to an adjustable accuracy $10^{-p}$. Recombinations, collisional 
ionizations and several other radiative processes which contribute  
to gas cooling, are instead reproduced as continuous by introducing 
the corresponding rate coefficients. For this reason errors on these
quantities are introduced by the discretization of the equations.
Hence the level of accuracy can be controlled very well if photoionization 
is dominant with respect to continuous processes by simply 
increasing $N_{cr}$; if $<\dot N_\gamma> (L_{box})^{-2} t_{min}$ becomes too low, 
the accuracy can instead be degraded: however, the evolution
of this quantity is monitored in order to prevent such degradation to occur 
during the simulation. \\
The computational cost of a given simulation is proportional to 
$N_c^3 N_{cr}= f_d N_c N_p N_s$, which gives  a rough estimate 
of the number of times the system eq.~\ref{eq15} is solved. 

\section{Tests}
\begin{figure}
\centerline{\psfig{figure=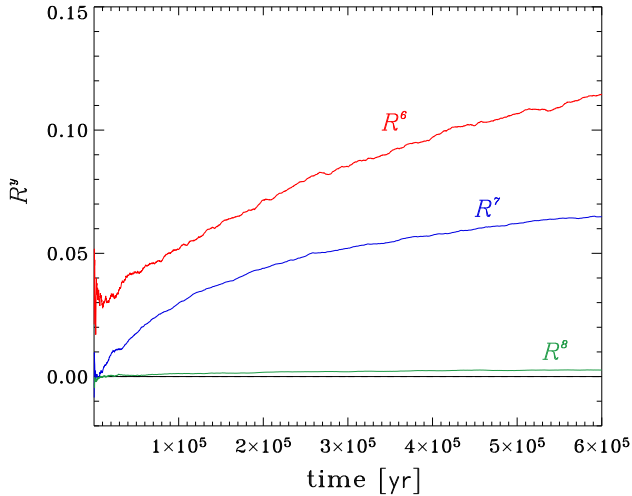,height=7cm}}
\caption{Time evolution of the residuals of the Str\"omgren sphere radius,
${\mathcal R}^y$, for a run with $N_p=10^y$ and $y$=6,7,8, calculated with
respect to $N_p=10^9$ (see text for details).} 
\label{fig1}
\end{figure}
In this Section,  we present a set of tests to illustrate the performances of \CR
in a wide range of astrophysical and cosmological contexts. 
Radiative transfer problems in three dimensions do not have, in general, analytical solutions 
except for some simple cases; for this reason testing quantitatively a radiative 
transfer code is a challenging purpose.
As a first test we consider the case of a monochromatic point source 
embedded in a pure-hydrogen homogeneous gas, with the ionized component
at constant temperature. This idealized case of an isothermal Str\"omgren sphere, 
has the advantage of being directly
comparable to an analytical solution of the problem ({\it e.g.} Spitzer 1978). 
Hence, it is possible to verify the accuracy of the code in reproducing the time 
evolution of the simulated system and its performances in a large 
range of gas densities. 
A second test is designed to check the reliability of the algorithm in
calculating the temperature and ionization structure of a gas of primordial
H/He composition.
\begin{figure}
\centerline{\psfig{figure=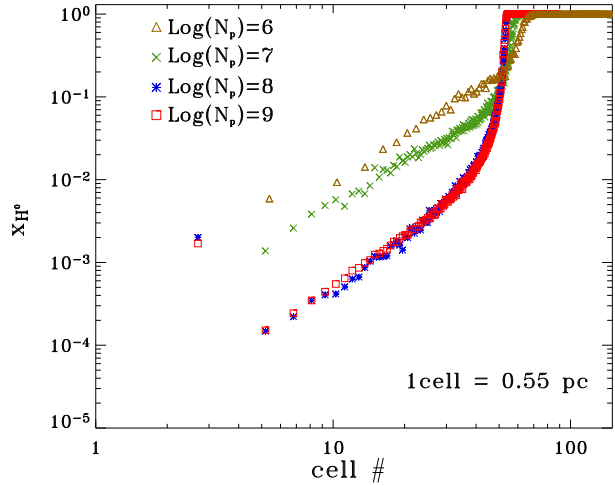,height=7cm}}
\caption{Neutral hydrogen fraction, $x_{{\rm H}^0}$, as a function of
the distance from the source in cell units. 
The points represent a subsample of grid points for 
runs with: $N_p=10^6,10^7,10^8,10^9$ (triangles, crosses, asterisks and squares).}
\label{fig2}
\end{figure}
As no analytical solution is available for this more realistic case, as a check,
we compare our results with the ones of the 1-D RT code
CLOUDY94\footnote{http://nimbus.pa.uky.edu/cloudy/}, available on the web. 
The above test is extended to the case of two point sources with largely different
luminosities. This problem turns out to be a difficult benchmark for RT codes (see Gnedin \& Abel 2001 
for a discussion).
Finally, we study the case of a dense, very optically thick slab illuminated by a 
power-law background radiation: our main aim is to show how \CR performs when confronted with 
the problem of the diffuse radiation field produced by recombining gas (shadowing).\\
To test even more complex problems (\ie time evolution, inhomogeneous density field)
it is instead necessary to compare results obtained with different RT codes; 
such an attempt (dubbed TSU$^3$) is currently under development\footnote{
http://www.arcetri.astro.it/science/cosmology/Tsu3/tsu3.html}.
\begin{figure*}
\centerline{\psfig{figure=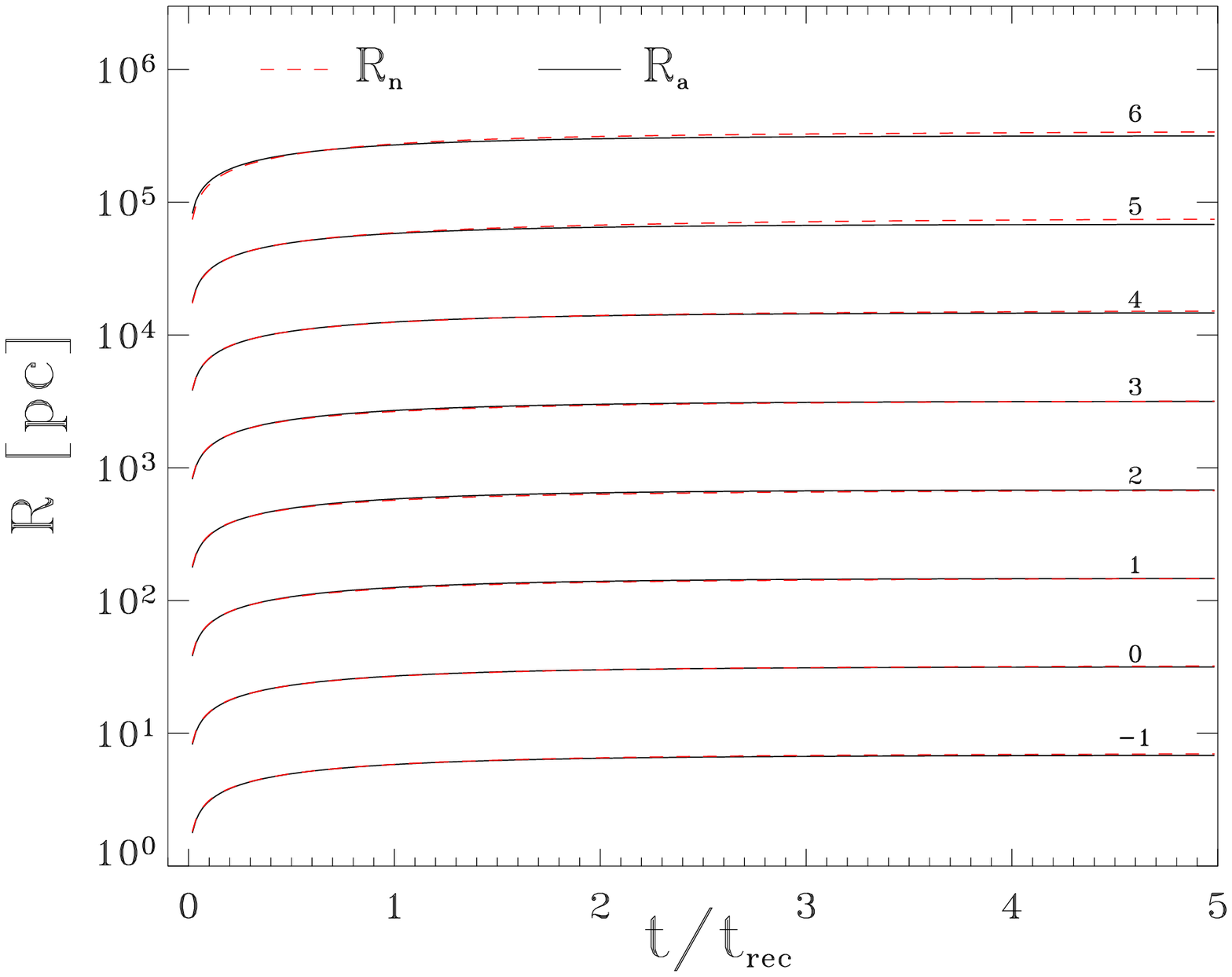,height=7.cm}\hspace{0.1cm}
\psfig{figure=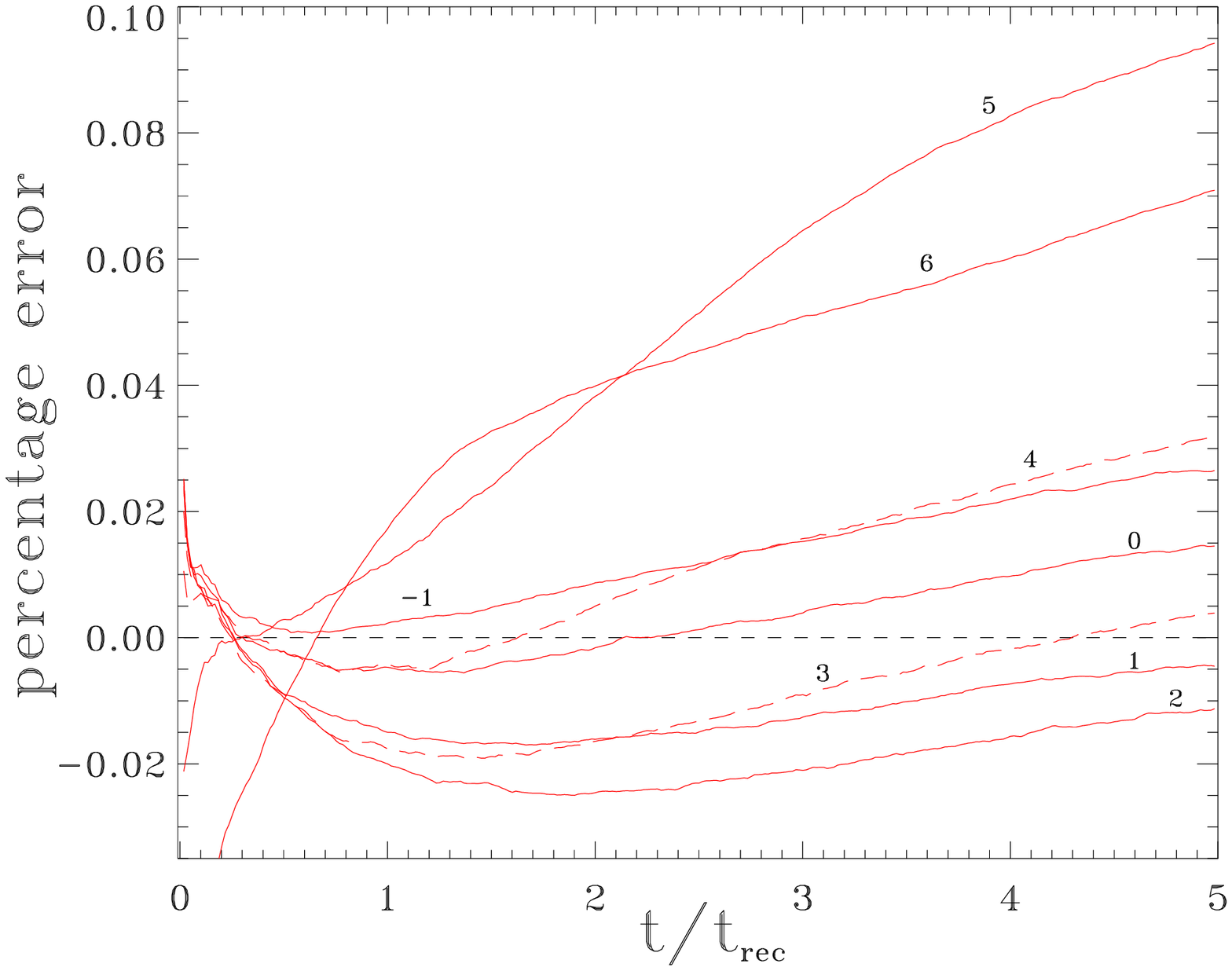,height=7.cm}}
\caption{{\it Left:} time evolution of the numerical 
H$_{\rm II}$ region radius $R_n$ (dashed lines) 
compared with the analytical solution $R_a$ (solid). 
The curves refer to eight different values of $n_H=10^{-q}$ cm$^{-3}$ 
and are labeled with the index, $q$,  of the corresponding density.
{\it Right:} percentage errors on 
the numerical radius with respect to the analytical one for the same runs 
(and labels) as in the left panel.}
\label{fig3}
\end{figure*}
\subsection{Pure Hydrogen, Isothermal Str\"omgren Sphere}        
As mentioned, we first test the code against the analytical solution for the time evolution 
of the radius of the H$_{\rm II}$ region produced by a monochromatic source with constant  
ionizing rate and expanding in a homogeneous medium. 
The comparison requires an helium abundance equal to zero and a constant temperature
in the H$_{\rm II}$ region; we fixed it equal to $T=10^4$ K. 
The (homogeneous) density field in the simulation box is 
initialized with a hydrogen number density of  $n_H=1$~cm$^{-3}$; the   
cartesian grid, which has a linear size $L_{box}=70$~pc,  is composed of $N_c^3=128^3$ cells. 
This density value has
been chosen in order for the diffuse (recombination) radiation effects to be evident, 
so to test the corresponding implementation.
The monochromatic source emits photons with energy equal to 13.6 eV, is located 
at the center of the box and has an ionizing photon rate $\dot N_{\gamma}=10^{48}$s$^{-1}$. 
The ionized hydrogen fraction is initialized at collisional equilibrium at $T=10^4$ K, \ie 
$x_{{\rm H}^+}=1.2\times 10^{-3}$.
To ensure that equilibrium configuration is achieved the
simulation is carried out for a physical time $t_s=6 \times 10^5$ yrs, 
roughly equal to five recombination times.
 
The H$_{\rm II}$ region equivalent radius, $R_n$, 
can be derived as $R_n=(3 V_n/4\pi)^{1/3}$, where $V_n$ is the volume of the ionized
region.
The contribution of each cell $(x,y,x)$ to $V_n$ is assumed to be a fraction of the cell 
volume equal to its ionization fraction $x_{\rm H^+}(x,y,z)$:
\be     
\label{eq20} 
V_n=\sum_{ix,iy,iz}x_{H^+}(\Delta x)^3.
\ee
As a single cell could be brought to ionization equilibrium in several time steps, 
this approach guarantees that the volume of the H$_{\rm II}$ region is not 
overestimated as the contribution of the partially ionized cells at the edge of the 
I-front is included. 

To check for convergency, we run four simulations with increasing $N_p$.
Fig.~\ref{fig1} shows the residual, ${\mathcal R}^y$, of a
given run with respect to the highest resolution run ($N_p=10^9$),
defined as:
\be
\label{eq21} 
{\mathcal R}^y=\frac{R_n^y-R_n^9}{R_n^9},
\ee
where $y=6,7,8$ refers to the run with $N_p=10^y$.
Good convergency is achieved  already for $N_p=10^7$, when  the residual is $\simlt 7$\%.\\
In Fig.~\ref{fig2}, the equilibrium neutral hydrogen fraction, $x_{{\rm H}^0}=n_{{\rm H}^0}/n_{\rm H}$,
profile is shown for the four runs.
In this case, a higher number of photon packets ($N_p \approx 10^8$) is needed to
reach a good level of convergency. 
This result suggests that applications which require the determination of the volume 
ionization structure (as cosmic reionization) 
are less demanding than problems that require the correct value of neutral hydrogen fraction 
(\ie Ly$\alpha$ forest simulations).
Thus, depending on the application and the desired accuracy, 
the required number of photon packets might vary; anyway \CR
converges for a reasonable number of photon packets.  
As $N_p=10^7$ already gives good convergency of the calculated numerical radius 
to be compared with the analytical solution, in the following applications we will
adopt this value, unless otherwise stated.

The time evolution of $R_n$ can now be compared
with the classical analytical solution (Spitzer, 1978):
\be
\label{eq22} 
R_a(t)=R_S (1-e^{-n_H \alpha_B t})^{1/3},
\ee
where $R_S=\left(3 \dot N_\gamma/4\pi n_H^2 \alpha_B \right)^{1/3}$ is the Str\"omgren 
radius and $\alpha_B$ is the hydrogen recombination 
coefficient to levels higher than the first. 
The comparison of our numerical results for $R_n$ against the analytical solution, 
has been performed for eight different values of the density in the range
$n=10^{-6} - 10$~cm$^{-3}$. In each run $N_p$, $N_c$ and  $\dot N_\gamma$ are
kept constant, as specified above. 
The source has been located at a corner of the simulation box (of linear size equal to
$6 R_S/5$), to maximize spatial resolution.
With the physical simulation time set to $t_s = 5 t^B_{rec}$, 
the equilibrium configuration is reached within $0.25$\% 
of the total computational time, according to eq. (\ref{eq22}). 
As $N_{cr}$ is fixed for all the runs, the accuracy itself is the same, as far as
the photon packet distribution and the discretization errors are concerned. 
Similarly, the ratio $N_\gamma/N_a$, which determines the accuracy to which 
the continuity of the photoionization process is reproduced, is the same for all 
the runs. In fact,
\be
\label{eq23} 
N_\gamma / N_a&=&(\dot N_\gamma t_s N_p^{-1})/(n_H {\Delta x}^3)\\
&\propto& n_H^{-1} / \left[ n_H (n_H^{-2/3})^3 \right]={\rm const}\nonumber,
\ee 
Finally, the ratio between the ionizing photon emission rate and the 
rate of physical processes reproduced as continuous (recombinations, 
collisional ionizations etc.) in a given cell increases with density: 
$\dot N_\gamma (L_{box})^{-2}t_{min} \propto \dot N_\gamma 
(n_H^{-2/3})^{-2} n_H^{-1}\propto n_H^{1/3}$. 
Hence, if  $\dot N_\gamma$ remains constant,  
the accuracy of the results decreases towards lower densities for the reasons 
explained in Section 2.5. 

This conclusion is supported by the 
numerical experiments shown in Fig.~\ref{fig3}, showing 
the evolution of the numerical radius, $R_n$ (dotted line), compared 
with the analytical solution, $R_a$ (solid line), 
for eight values of the density considered. 
The agreement is remarkably good. For densities $ > 10^{-4}$cm$^{-3}$ 
errors remain through the entire simulation $\simlt 3$\%; 
for lower densities they increase, according to expectations following
the previous discussion, up to $\approx 10$\% for $n_H=10^{-5}- 10^{-6}$ cm$^{-3}$.  
We point out that these runs have the minimum resolution required to
achieve convergency and that accuracy can be easily improved using a larger 
value for $N_p$.

These tests, embracing 8 dex in density, demonstrate that \CR is quite
robust and accurate over a wide range of physical conditions.
\begin{figure*}
\centerline{\psfig{figure=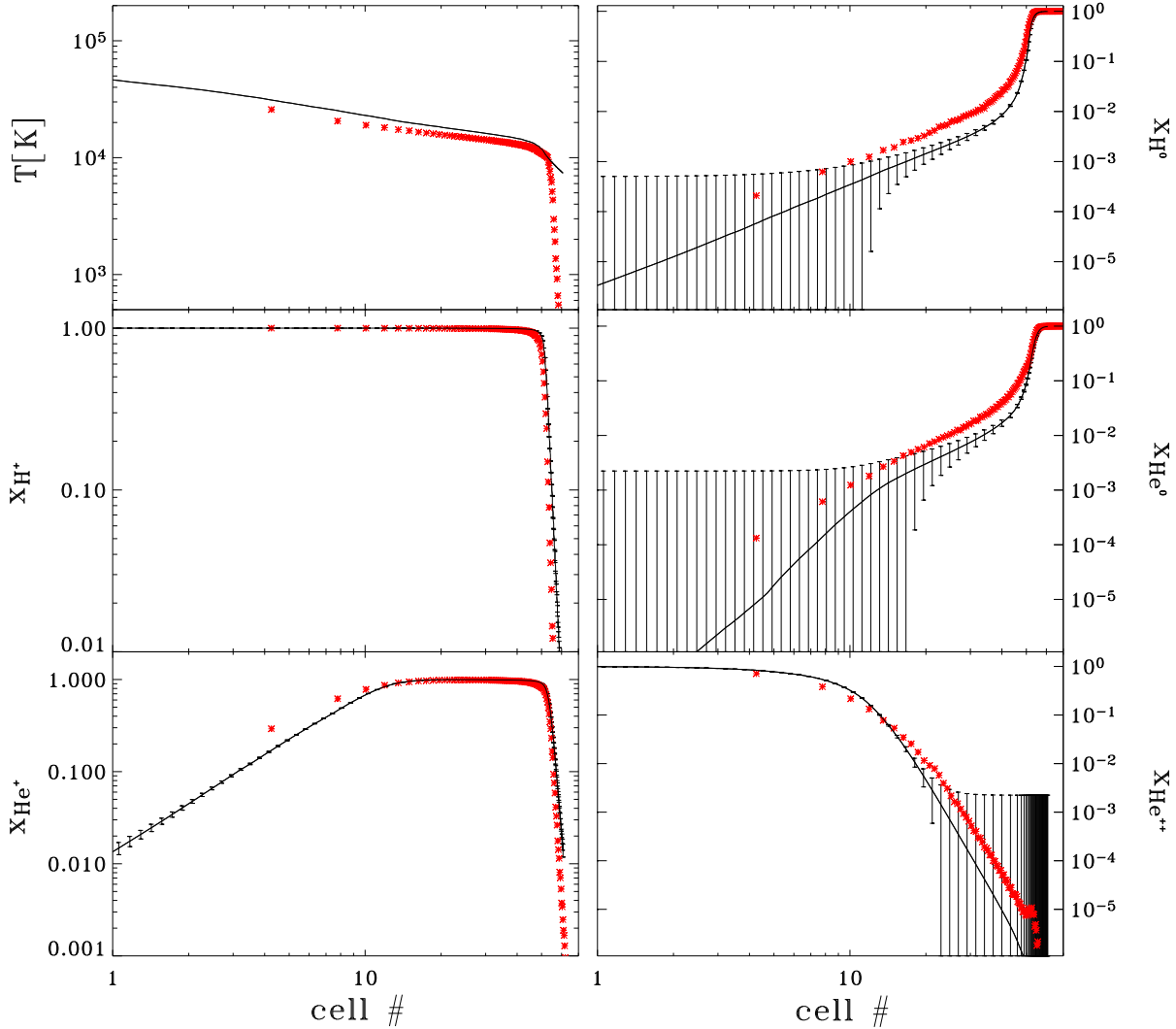,height=13cm}}
\vskip 1truecm
\caption{
Comparison between the equilibrium configurations obtained by 
\CR (points) and CLOUDY94 (solid lines)
for different physical quantities, 
as a function of distance from the point source in cell units. 
The errors refer to the CLOUDY94 results and are evaluated
as explained in the text. }
\label{fig4}
\end{figure*}
\subsection{Realistic Str\"omgren Spheres}
Next, we test the \CR scheme for the calculation 
of temperature and helium ionization. As no analytical solution is
available, we compare the results with those obtained with the public code
CLOUDY94, a 1-D radiative transfer code which calculates {\it equilibrium}
configurations of the various species. The physics included in the two codes is
slightly different. CLOUDY94 treats
atoms and ions as multilevel systems, while we consider all species as one-level
systems; furthermore CLOUDY94 includes some physical processes, like heat conduction and
gas pressure effects, which are not included in \CR.
We consider a point source, emitting as a black body at $T=60000$ K with a 
luminosity $L=10^{38}$~erg~s$^{-1}$, located at the center of the 
simulation box and embedded in a homogeneous density field with $n=1$ cm$^{-3}$ 
composed of hydrogen (90 \% by number)  and helium (10~\%). 
The gas is initially completely neutral and at a temperature  $T=10^2$ K 
in the entire simulation box, whose linear dimension has been fixed to $L_{box}=128$ pc. 
As an accurate determination of the neutral fractions and temperature inside the
ionized regions requires $N_p \ge 10^8$ (see previous
Section), we adopt $N_p =10^8$.
Moreover, spherical symmetry is required to
degenerate our 3-D solution to the 1-D geometry adopted by CLOUDY94. 
The comparison is performed at a time $t_s=6 \times 10^5$~yr $\approx 5 t_{rec}^B$
(where $t_{rec}^B$ is the characteristic time scale for hydrogen 
recombination to levels other than the first), 
when the equilibrium configuration has
been reached.
In Fig.~\ref{fig4} the comparison between the results of the CLOUDY94
(solid lines) and {\tt CRASH} (points) is shown. The value of the different
physical quantities is plotted as a function of the distance from the source, 
expressed in cell units, $\Delta x= 1$ pc.
The points represent spherically averaged \CR outputs. 
Despite the differences between the two codes, 
a remarkably good agreement is obtained.
In particular, the temperature profiles agrees within 10\% 
of the CLOUDY solution, 
except from the warm tail
extending beyond the location of the ionization front produced by CLOUDY94. 
The nature of this feature is unclear, but it can be probably imputed   
to heat conduction, implemented in CLOUDY94,  
which cause an heat transfer across the ionization front where the temperature 
gradient is very high.
\begin{figure*}
\centerline{\psfig{figure=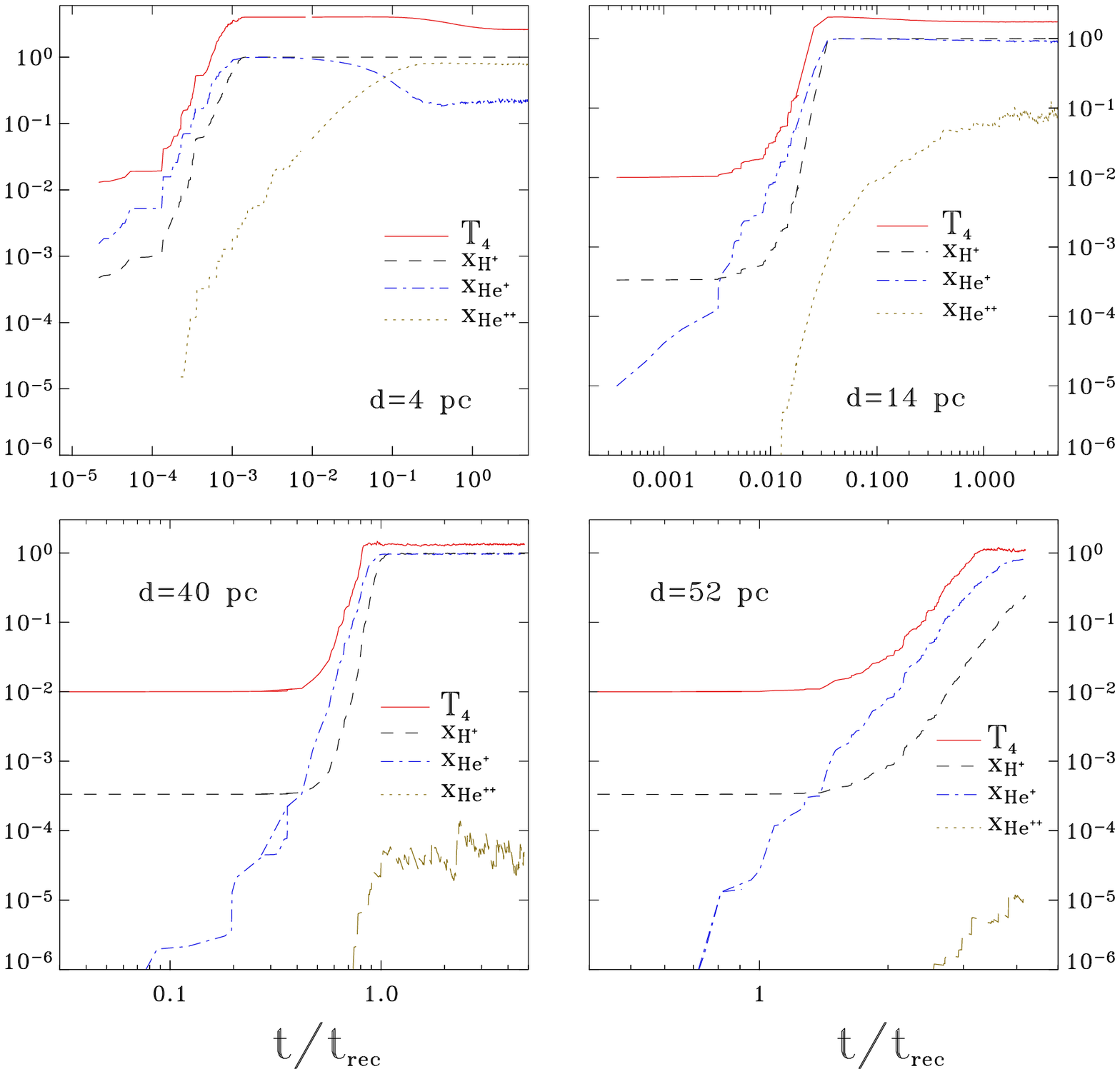,height=13cm}}
\caption{Evolution 
of the temperature, $T$ (in units of $10^4$ K; solid lines) 
and ionization fractions for $x_{H^+}$ (dashed), 
$x_{He^+}$ (dotted-dashed) and $x_{He^{++}}$ (long dashed), 
in four fixed cells at different distances from the source, 
$d$=1, 14, 40 and 52~pc. }
\label{fig5}
\end{figure*}
As for the ionization fractions, the errors plotted on the CLOUDY94     
curves are evaluated as follows. At some distances from the source, the CLOUDY94     
calculation gives the unphysical results $(x_{H^+}+x_{H^0})> 1$ 
or $(x_{He^0}+x_{He^+}+x_{He^{++}}) > 1$; we then assume the errors to be 
the maximum value of the quantities  $\left[(x_{H^+}+x_{H^0})-1\right]$, 
for hydrogen species, and $\left[(x_{He^0}+x_{He^+}+x_{He^{++}})-1\right]$ 
for helium species. 
The errors on the \CR results are shown by the scatter of the plotted points, which is more
significant for the neutral fractions as a result of  the discretization error connected to 
continuous process. 
The $x_{H^0}$ and $x_{He^0}$ profiles produced by the two codes are 
characterized by the same trend. The \CR results tend to overpredict these quantities
with respect to CLOUDY94 on average by 20-30\%; 
this is also the case for the $x_{He^+}$, and is likely due to differences in 
the expressions for the rate coefficients adopted. 
For example, at $T=10^4$ K the \CR total hydrogen recombination rate (see Appendix B)
is equal to 5.097$\times 10^{-13}$ cm$^{3}$ s$^{-1}$,  while that adopted by CLOUDY 
(see Table 20 of CLOUDY90 manual) is 4.17$\times 10^{-13}$ cm$^{3}$ s$^{-1}$, 
\ie a 20\% discrepancy.
Our higher value is consistent with the larger abundance of H$^0$ found inside the H$_{\rm II}$.
The $x_{H^+}$ profiles are in perfect agreement, 
as well as the $x_{He^{++}}$ profiles which appear 
to be in very good agreement within the errors. 

To get  more insights on the numerical technique adopted, we plot in Fig.~\ref{fig5} the evolution 
of the temperature and ionization fractions for the various species
in four fixed cells at different distances from the source 
($d= 4, 14, 40$ and 52~pc). 
It is important to note that the continuity of the photoionization 
(photoheating) process is well reproduced 
at all distances; also, cells at different distances start to be ionized 
at different times, as the I-front propagates into the density field. 
Only in the cell closest to the source the radiation field at energies 
higher than 4~Ryd is strong enough to significantly doubly ionize He; at 
14 pc from the source He$^{+}$ is only partially ionized, and in the
two most distant cells He$^{++}$ aboundance is almost negligible. 
\begin{figure*}
\centerline{\psfig{figure=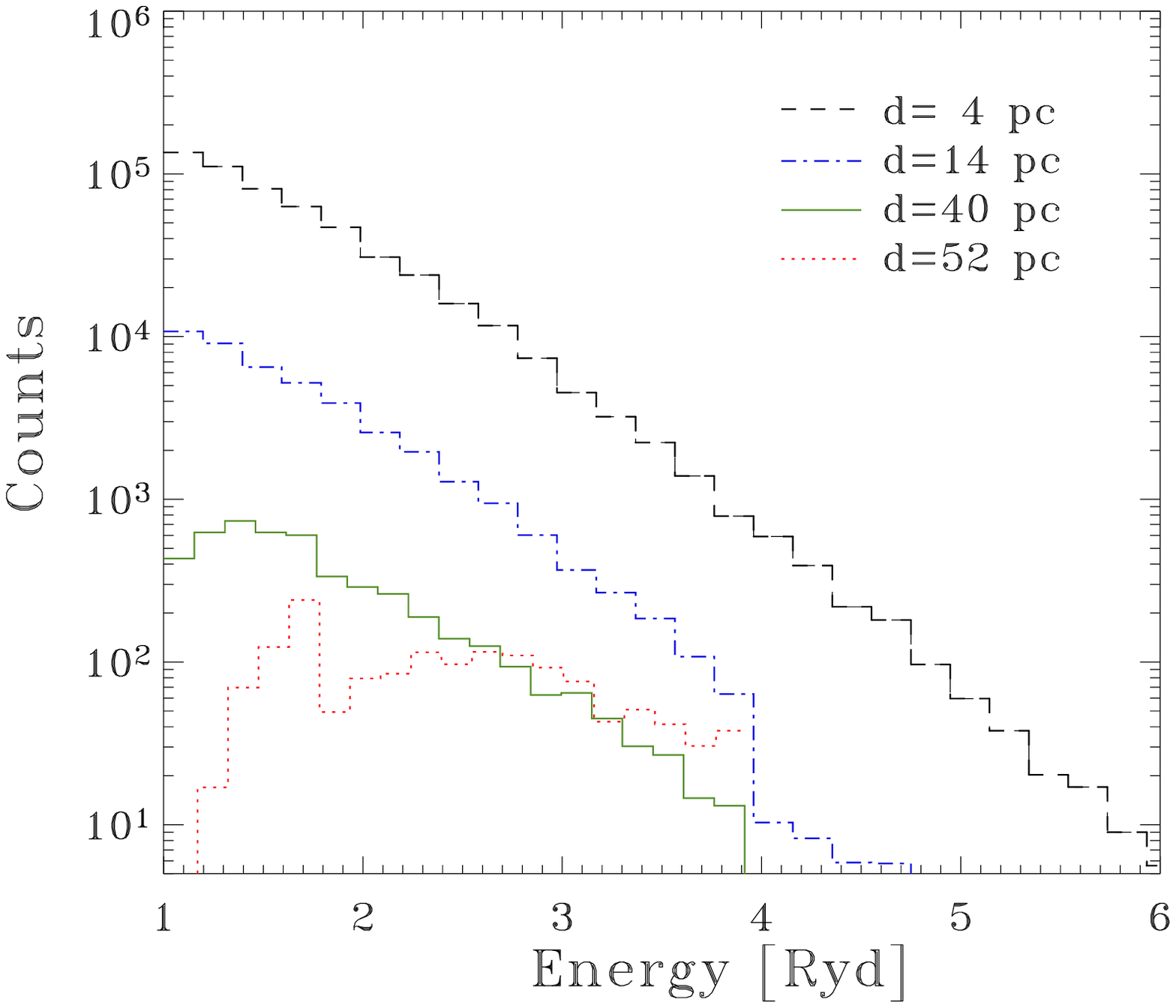,height=7cm}}
\caption{Frequency distribution of the photon packets, weighted by their photon 
content,  
seen by the same four cells of Fig.~\ref{fig5} over the entire
simulation. }
\label{fig6}
\end{figure*}
The flux depletion and the progressive filtering 
of the spectrum by the intervening medium,  
are further illustrated by Fig.~\ref{fig6}, where we plot 
the frequency distribution of photon packets crossing the same four cells 
as above during the entire simulation, weighted by their 
photon content.  
Moving to regions farther from the source, the ionizing flux 
decreases, as expected, due to the combined effect of geometrical dilution 
and photoelectric absorption; this last process also causes the filtering 
effect resulting in the progressive hardening of the spectrum.
Again, we note that packets with energy above 4 Ryd are heavily 
depleted in distant regions.

These tests allow us to conclude that \CR reliably computes, on limited computational times 
(the discussed simulation took a few hours on a 1 GHz workstation), 
the physical quantities of a photoionized gas of primordial composition
and, when confronted to state-of-the-art 1-D equilibrium codes as
CLOUDY94, produces results of comparable accuracy level.  

As a final remark, we note that the presented tests constitute  
a particularly difficult physical benchmark due to the relatively  
high densities and soft source spectrum. 
\begin{figure*}
\centerline{\psfig{figure=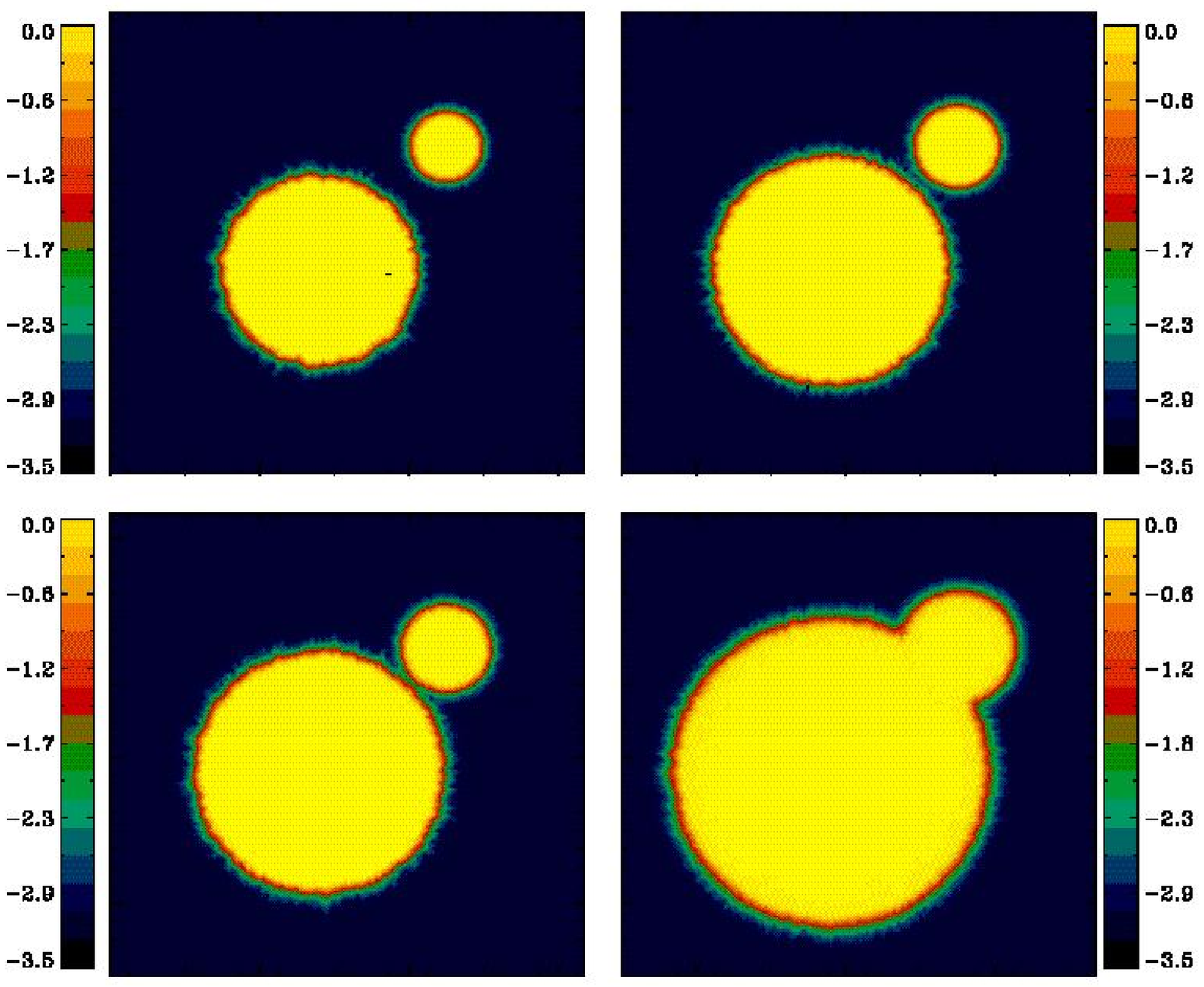,height=12cm}}
\caption{Slice of the simulation box through the planes
containing the source locations showing the ionized hydrogen fraction, $x_{\rm H^+}$,  
distribution at four different 
simulation times: $(0.3, 0.6, 0.9, 3) \times t_{rec}^B$ (from left to 
right and top to bottom).}
\label{fig7}
\end{figure*}
\subsection{Multiple Point Sources}
As a third test we run the case of multiple radiation sources in the box.
Again, we consider a homogeneous H/He density field with $n=1$ cm$^{-3}$,
$N_c=128$, $N_p=10^8$ and $L_{box}=160$~pc. 
Two sources are embedded in the computational volume: one 
with $L_1=10^{38}$ erg s$^{-1}$, and  a weaker one 
with $L_2=L_1/27$, both with a $T=60000$~K black-body spectrum. 
The temperature and ionization fractions fields are initialized as in the 
previous test. 
We follow the evolution 
of this system for  a time $t_s=4.5 \times 10^5$~yr $\approx 3.4 t_{rec}^B$. 

In Fig.~\ref{fig7}  we show the ionized hydrogen fraction distribution in a slice of the 
simulation box through the plane containing the source locations. The panels refer to
four different simulation times: $(0.3, 0.6, 0.9, 3) \times t_{rec}^B$.
The H$_{\rm II}$ regions around each source evolve 
independently until they merge at $t\simeq t_{rec}^B$. Once the H$_{\rm II}$ regions
overlap,  photons from one source can change the properties of the
ionized region previously produced by the other one.
In particular, we notice that the I-front curvature of the smaller H$_{\rm II}$
region is changed in the overlapping region. 
\begin{figure*}
 \centerline{\psfig{figure=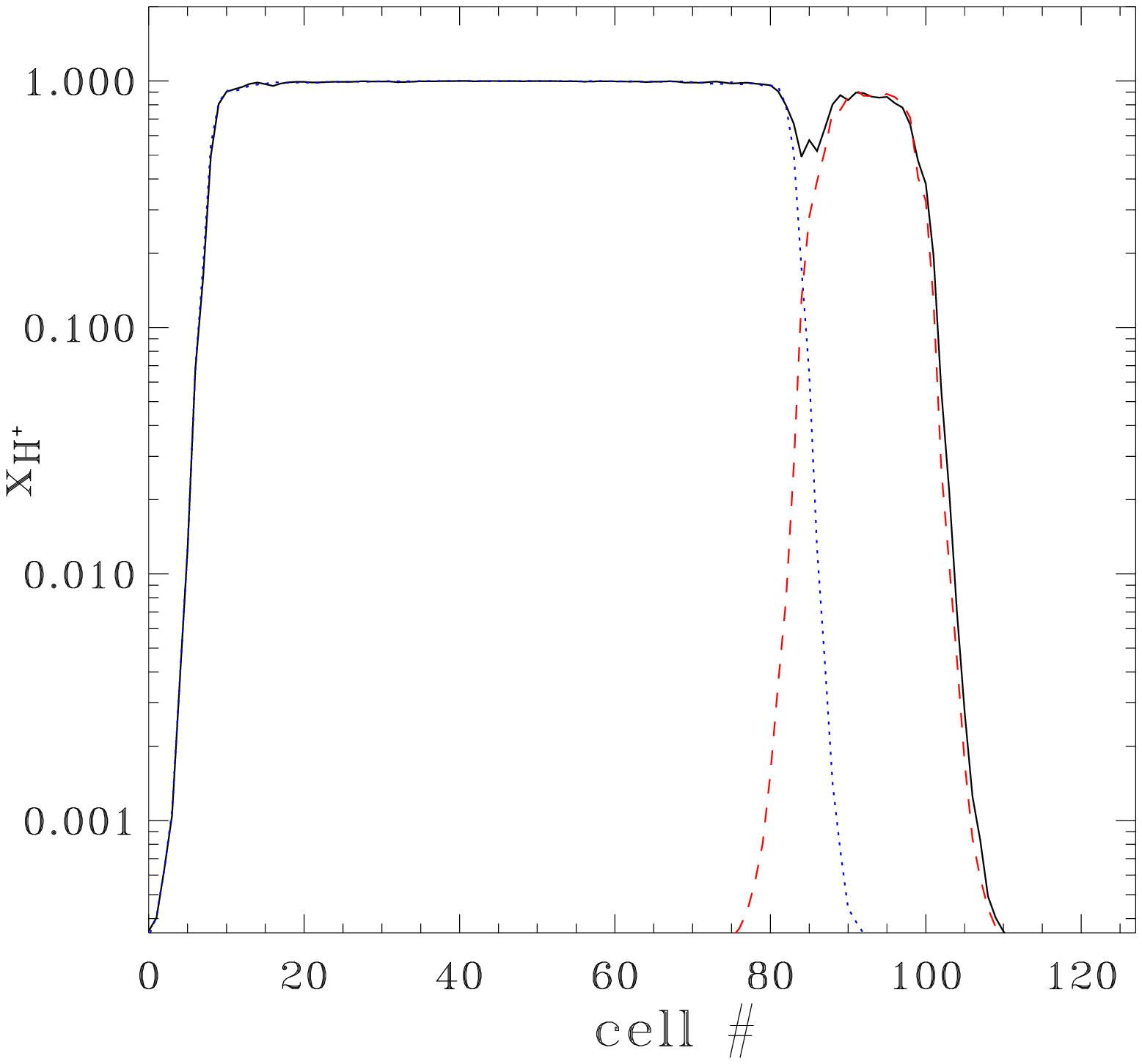,height=8cm}\hspace{0.1cm}
 \psfig{figure=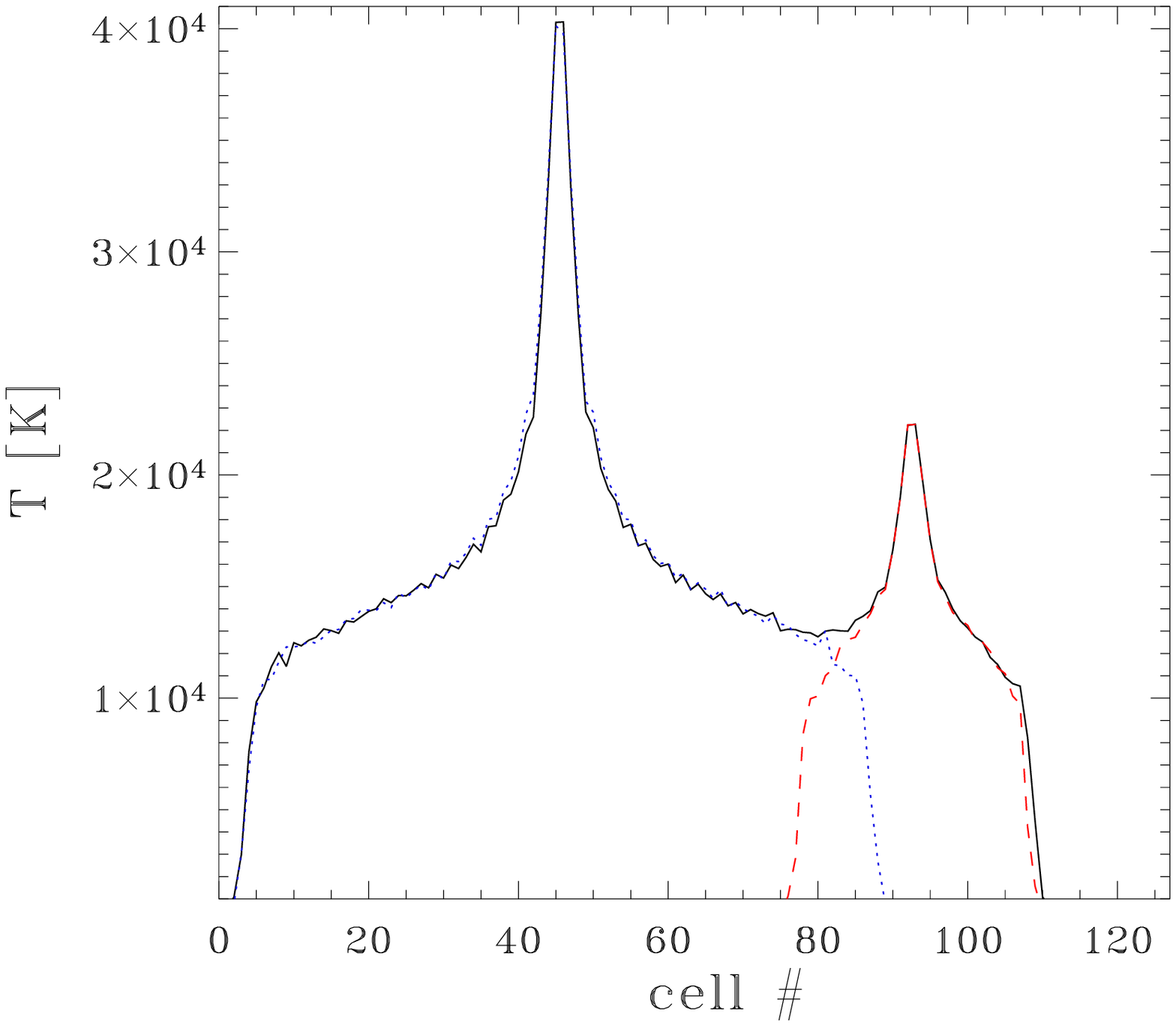,height=8cm}}
 \caption{{\it Left:} Ionized hydrogen 
fraction along a direction parallel ($\approx 10$ cells away) 
to the one connecting the two sources and passing trough the 
I-fronts intersection region. The three lines refer to simulations in which
either the stronger (dashed line) or the weaker source (dotted-dashed) were turned off; 
the third cut (solid) is through the simulations in which both sources are active. {\it Right:} 
Temperature along the line connecting the two sources; the notation for the curves is 
the same as in the left panel.}
\label{fig8}
\end{figure*}
\begin{figure*}
\centerline{\psfig{figure=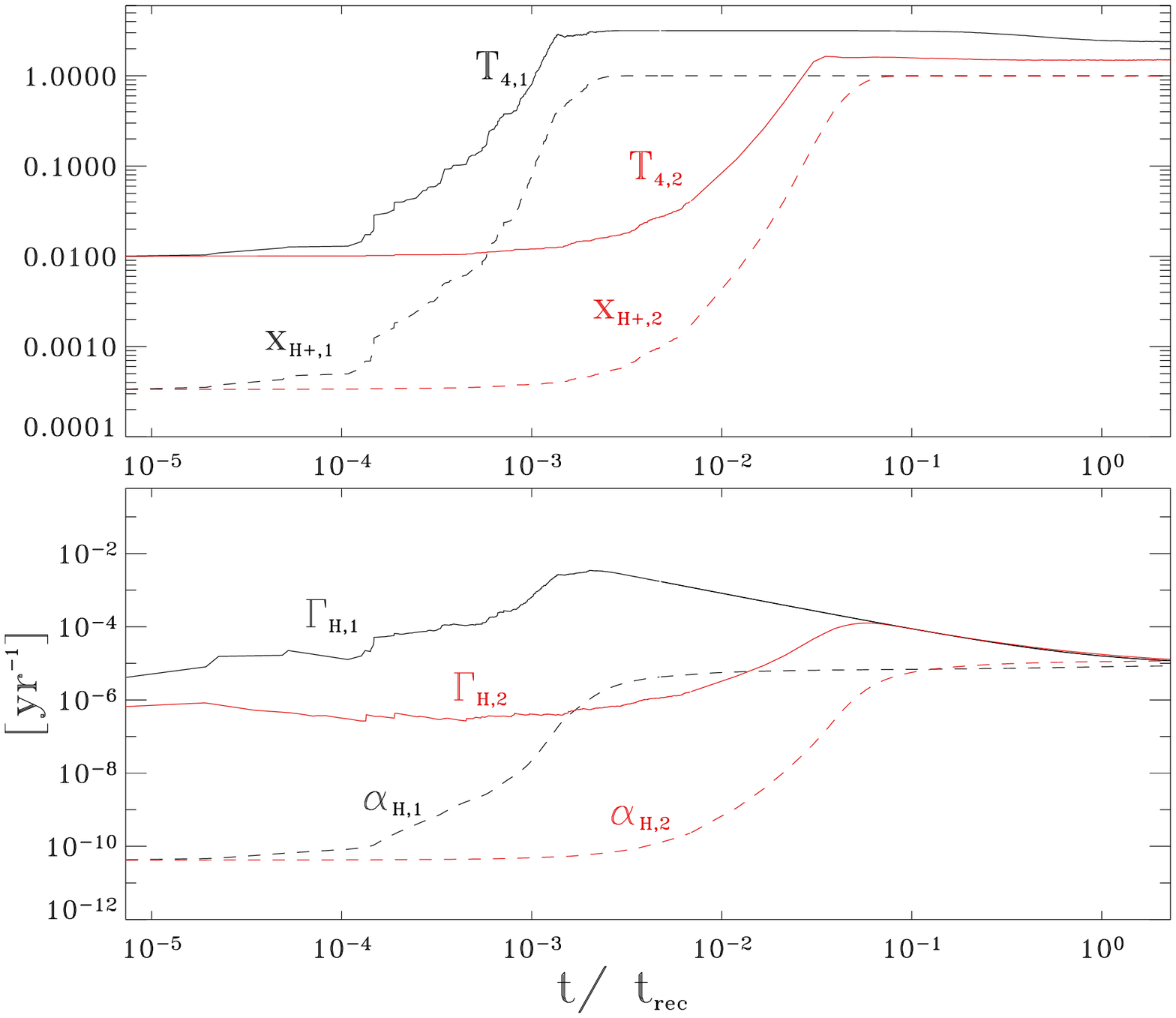,height=12cm}}
\caption{Upper panel: Time evolution of the 
temperature (solid), hydrogen ionization fraction (dashed) for two cells
located at a distance $d=5$~pc from each source. The label 1 (2) 
on the curves refer to the cell at 5 pc from the the source wiht luminosity 
$L_1$ ($L_2$=$L_1/27$). Lower panel: As above, but for the 
photoionization (solid) and recombination (dashed) rates. 
}
\label{fig9}
\end{figure*}
This effect is better illustrated by Fig.~\ref{fig8} (left panel), where
a 1-D cut of the ionized hydrogen fraction along a direction parallel ($\approx 10$ 
cells away) to that connecting two sources is shown.
The three lines refer to simulations in which either one or the other source were 
turned off and in which both sources are active. As described above, in the cumulative curve
an enhancement of the H$^+$ fraction is clearly visible both in the overlapping region
and on the far side of the smaller H$_{\rm II}$ region due to
percolating photons coming from the stronger source.  
A similar effect is visible in the temperature profiles along the line of 
sight connecting the two sources, plotted in the right panel of Fig.~\ref{fig8}. 
In the overlapping region the temperature is larger than for the 
case of a single source.

In order to understand the reason for the difference in the temperature profiles 
within the H$_{\rm II}$ regions of the two sources we have performed an additional check. 
The heating rate associated with photoionizations of the species $A$ can be written as  
${\cal H}_A = \Gamma_A \langle h \nu \rangle$, where $\langle h \nu \rangle$ is the mean 
energy of the photoelectrons. As the spectrum is the same, ${\cal H}_A$ only depends on the 
photoionization rate. The temperature results from the balance of such photoheating
with radiative losses. In turn, we have checked that the latter are largely dominated by 
the recombination cooling rate, which can also be written as the product of the 
recombination rate and the 
mean energy lost per recombining electron. Fig.~\ref{fig9} shows   the time evolution of 
the temperature, hydrogen ionization fraction (upper panel), photoionization 
and recombination rates (lower panel) for two cells located at a distance $d=5$~pc 
from each source. The different luminosity of the sources results in a different 
gas thermal/ionization history, the cell illuminated by the most luminous source 
reaching equilibrium earlier.
The photoionization rate increases smoothly with time, due to the decrease of the
intervening opacity, until it reaches a maximum;  thereafter it decreases because 
of the lack of absorbers. 
The temperature equilibrium value is reached in both cases slightly before the 
recombination rate reaches a 
plateau at $\approx 10^{-5}$~yr$^{-1}$; this value is the same for the two sources as
it essentially depends only on the gas density (the tiny deviation seen  at large
evolutionary times is due to the temperature dependence of the recombination 
coefficient). However, the photoionization rate at that
stage is larger (by roughly a factor $L_1/L_2$=27) for the more luminous source 
than for the fainter one. This
extra net heating explains the temperature difference. It is also worth noting
that asymptotically the photoionization and the recombination rates become
equal as expected from the equilibrium condition for a Str\"omgren sphere.

\begin{figure*}
\centerline{\psfig{figure=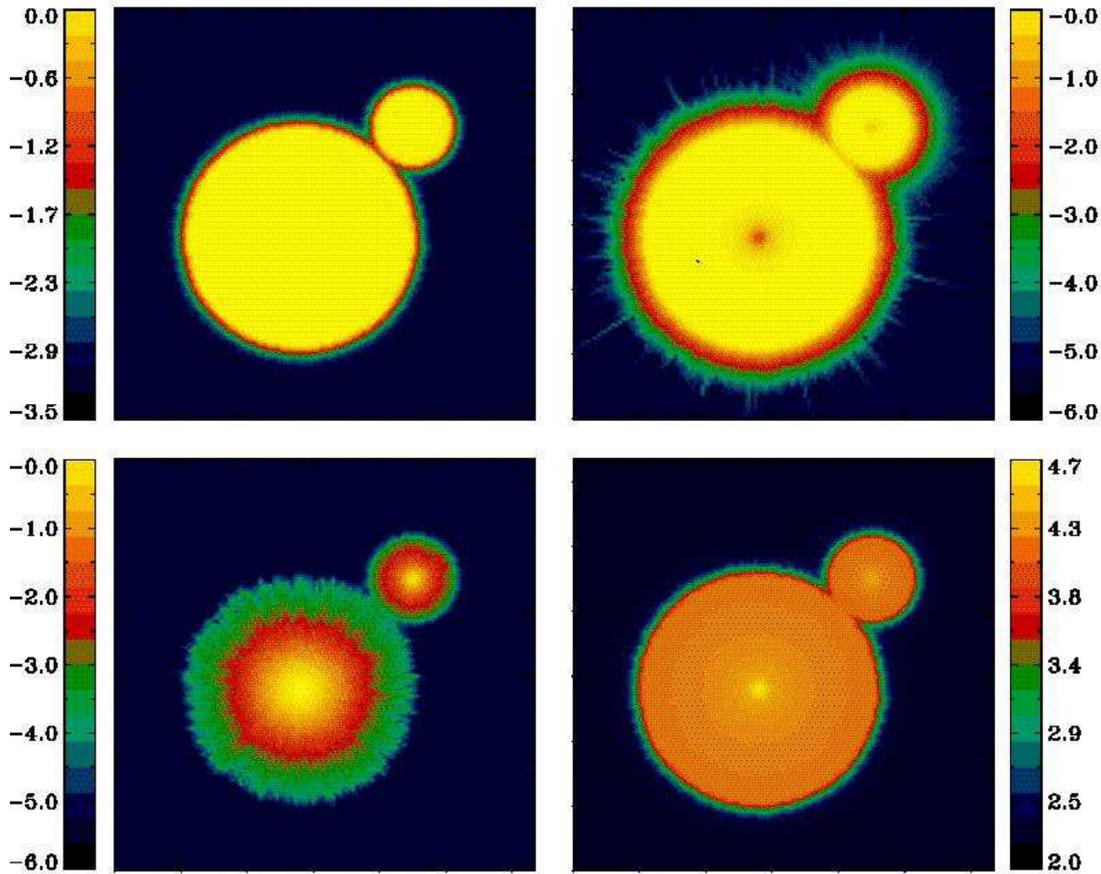,height=12cm}}
\caption{Slices of the simulation box through the plane
containing the sources locations showing  $x_{H^+}$, $x_{He^+}$,
$x_{He{^{++}}}$ and temperature distributions (from left to
right and from top to bottom) at 
the end of the simulation, $t=3.4 t_{rec}^B$.}
\label{fig10}
\end{figure*}
Finally, in Fig.~\ref{fig10} we plot 2-D maps of  the 
$x_{H^+}$, $x_{He^+}$, 
 $x_{He^{++}}$ and temperature distributions at the end of the simulation. 
Due to the larger mean free path typical of He$^0$ ionizing photons, the 
He$^+$ I-front is  wider than the H$^+$ one, and it is also smoother 
due the flatter spectrum at $h\nu \ge 24.6$ eV. 
Moreover, the He$^+$ I-front appears to be more affected by numerical noise 
because of the poorer sampling of the helium ionizing region of the spectrum. 
This is even more evident in the He$^{++}$ distribution for which the I-front 
is jagged in the outer parts, where ionization balance is governed 
by the very high energy tail of the power spectrum. Nevertheless 
the calculation correctly reproduces the higher He$^+$ ionization level 
around to the stronger source.
\begin{figure*}
\centerline{\psfig{figure=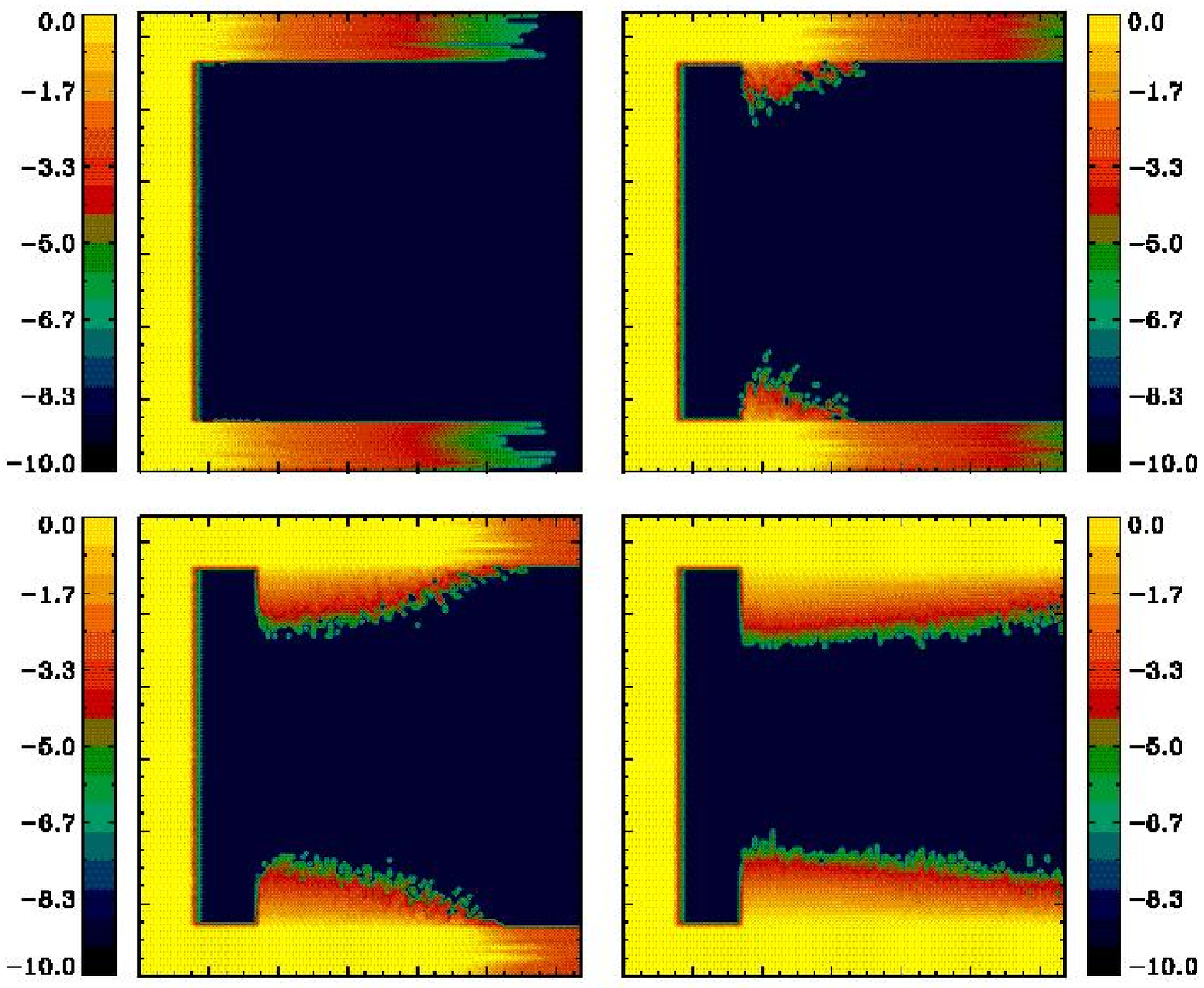,height=12cm}}
\caption{Maps of the $x_{H^+}$ distribution at four different simulation times: 
$(0.2, 0.4, 0.9, 2)\times t_{rec}^B$ (from top to bottom and from left to right).
The test refers to a region with an overdensity of $10^3$ and a dimension of 
128$\times$100$\times$20 cells embedded in a homogeneous medium and exposed to a plane
parallel ionization front traveling from left to right (see text for details).}
\label{fig11}
\end{figure*}
\begin{figure*}
\centerline{\psfig{figure=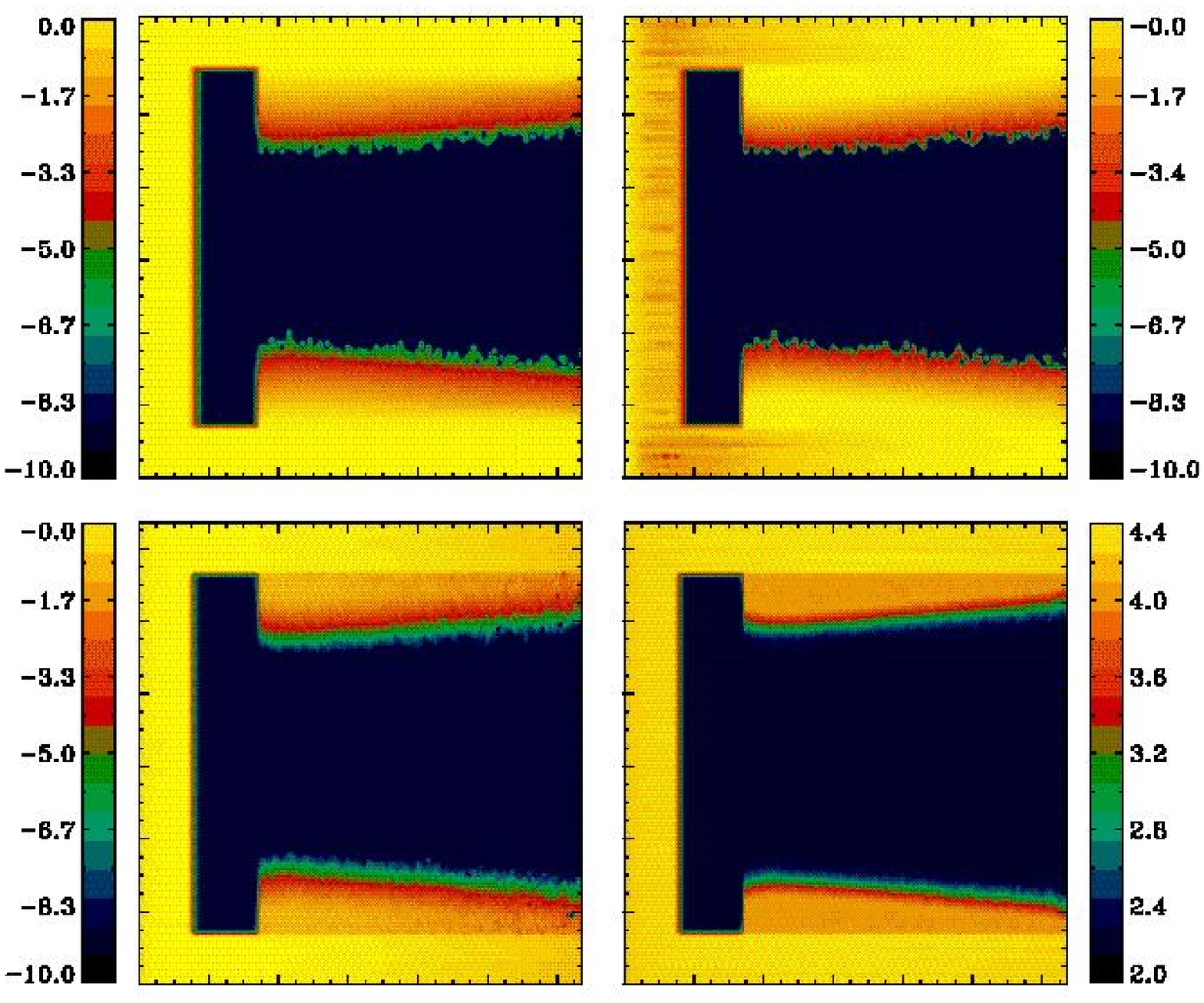,height=12cm}}
\caption{Maps of $x_{H^+}$, $x_{He^+}$, $x_{He{^{++}}}$
and temperature distributions (from left to right and from top to bottom) at
 the end of the simulation.}
\label{fig12}
\end{figure*}
\subsection{Shadowing of background radiation}

As a final test, we study the  case of a gas ionized by a power-law background  
radiation field. Our main aim is to demonstrate the ability of \CR to deal
with the problem of diffuse recombination radiation. The density field has 
been initialized with a high density region ($n$=1~cm$^{-3}$) of dimension
128$\times$100$\times$20 cells, embedded in an  homogeneous 
gas with  $n=10^{-3}$ cm $^{-3}$. 
The density field is exposed to a plane parallel ionization front traveling 
orthogonally to the overdense slice.  The background spectrum has a power-law shape,
$J_\nu=J_0 (\nu/\nu_{th,H^0})^{-\alpha}$, with $J_0=10^{-22}$ erg cm$^{-2}$ s$^{-1}$ Hz$^{-1}$ 
and $\alpha=-1.4$. The gas composition is the same as in the previous tests. 
The gas is initially completely neutral and at a temperature  $T=10^2$ K 
in the entire simulation box; 
the physical simulation time is $t_s=3 \times 10^8$~yr$\approx 4.3t_{rec}^B$, 
the box linear size is $L_{box}=6.6$~kpc; moreover, $N_p=5\times 10^7$ and $N_c=128$. \\
Fig.~\ref{fig10}  shows maps of the ionized hydrogen fraction distribution, 
orthogonal to the overdense slice, at four different simulation times: 
$(0.2, 0.4, 0.9, 2)\times t_{rec}^B$. 
The I-front initially propagates into the low density field, but 
it is stopped at the edge of the slice by the high recombination rate, producing
a shadow behind it.
The I-front is very smooth due the power-law spectrum of the ionizing background. 
As the simulation proceeds, the diffuse radiation produced 
by recombinations in the ionized gas, begin to ionize the low 
density region behind the high density slice.  
The diffuse radiation then propagates into the shadowed region, 
ionizing gas that would otherwise remain neutral.  
Finally,  Fig.~\ref{fig11}  shows maps for the final 
$x_{H^+}$, $x_{He^+}$, $x_{He^{++}}$ and temperature distributions.
The hard spectrum is able to ionize almost completely helium to He$^{++}$ 
in the regions exposed to direct radiation, whereas the diffuse radiation 
is not hard enough to produce the same effect. In the He$^{++}$ 
distribution (lower-left panel), a clear separation between the gas ionized by 
either direct or diffuse radiation is visible. 
A similar signature is present also in the temperature map, again caused by  
the fact that diffuse photons have typical energies confined in a narrow band 
above ionization thresholds of the three absorbers; as a consequence, little energy
is available to photoheat the gas.
In conclusion, although necessarily qualitative due to the lack of precise solutions to this
problem, the shadowing experiment performed here seems to yield physically
plausible results.
 
\section{Summary}
We have presented an updated version of \CR,
a 3-D RT code evaluating the effects 
of an ionizing radiation field propagating through a given       
inhomogeneous H/He density field on the physical 
conditions of the gas.  
The code, based on a Monte Carlo technique, self-consistently calculates the 
time evolution of gas temperature and ionization fractions
due to an arbitrary number of point/extended sources and/or 
diffuse background radiation with given spectra. 
In addition, the effects of diffuse ionizing radiation following
recombinations of ionized atoms have been included.  
The code has been primarily developed to study a number of 
cosmological problems, such as hydrogen and helium reionization,  
physical state of the Ly$\alpha$ forest, escape fraction of 
Lyman continuum photons from galaxies, diffuse Ly$\alpha$ emission
from recombining gas, etc. However, its flexibility allows applications that 
could be relevant to a wide range of astrophysical problems. 
The code architecture is sufficiently simple that additional physics
can be easily added using the same algorithms already described in
this paper. For example, dust absorption/re-emission can be included
with minimum effort; molecular opacity and line emission, although
more complicated, do not represent a particular challenge given the
numerical scheme adopted. Obviously, were such processes added, the 
computational time could become so long that parallelization would
be necessary. This would be required also when {\tt CRASH} will be coupled
to a hydrodynamical code to study   
the feedback of photo-processes onto the (thermo-)dynamics
of the system. This perspective development looks quite encouraging:
in fact, Monte Carlo schemes are by construction 
extremely suitable for parallel computations, as the packet load can
be distributed in a straightforward manner to processors, limiting the 
inter-communication problems among them to a very minimum. Clearly,
this is a noticeable advantage of the adopted scheme 
over those based on full solution of the RT equations.  
This study is already under work and its implementation
will be presented in a forthcoming communication. 

To demonstrate the performances, accuracy and robustness of the code 
we have studied in detail four different test cases designed to ascertain
specific aspects of RT: (i) pure hydrogen isothermal Str\"omgren spheres; 
(ii) realistic Str\"omgren spheres; (iii) multiple
overlapping point sources, and (iv) shadowing of background radiation. 
When possible, a detailed quantitative comparison of the results against 
either analytical solutions or 1-D standard photoionization codes
has been made showing a remarkable level of agreement ($\simlt$ few percent). 
For more challenging tests the code yields physically plausible results,
which could be checked only by comparison with other similar codes.\\

This work was partially supported by the Research and Training Network
`The Physics of the Intergalactic Medium' set up by the European Community
under the contract HPRN-CT2000-00126 RG29185.

\appendix
\section{Monte Carlo sampling}
The Monte Carlo (MC) method, extensively used in many field of scientific research, 
allows to sample with great efficiency physical quantities with a given Probability 
Distribution Function (PDF). Here we outline the basic principles on which the 
MC sampling technique is based. 
Let us consider a variable $q$ having values in the interval $[a,b]$, representing   
a physical quantity distributed according to a 
PDF, $f(q)$, normalized to $\int_a^b f(q)dq=1$. The PDF is defined so that
the probability of having a value $q$ in the range $[q',q'+dq]$ is 
$f(q')dq$. Let us also introduce the 
cumulative probability $P(q')$, defined as: 
\be
\label{A1}
P(q')=\int_a^{q'}f(q) dq, 
\ee
which measures the probability that the previous value of $q$ will yield
a value lower or equal to $q'$. Given the above normalization,
$P(q') \le 1$ for each $q'$ in $[a,b]$.

Once the PDF is given, a number ${\cal R}$ is randomly extracted in the
interval $[0,1]$. Then, $q'$ is derived setting ${\cal R}=P(q')$ and
inverting the integral in eq.~\ref{A1}.
In this way the probability of obtaining the value $q'$ when extracting randomly a
number in [0,1], corresponds exactly to $f(q')$. 

If the number of extractions is large enough, it is possible to obtain a statistical 
population of values which reproduces with great accuracy the given PDF. 
The main advantage of this technique lies in the ability to sample 
arbitrary distribution functions by simply randomly sampling the interval $[0,1]$, 
an easy and cheap algorithm to implement.
\section{Rate Coefficients and Cross Sections}
In this Section we list the rate coefficients 
and the cross sections adopted by \CR, for all the physical processes included.  \\

\begin{itemize}
\item{\it Photoionization cross sections} $\;\;\;$[cm$^{-2}$] $\;\;\;\;$(1)\\
$$
{\rm H}^0\;:  \;
\sigma_{H^0}(\nu)=6.3\times 10^{-18}(\nu/\nu_{th,{\rm H}^0})^{-3}
\;\;\;\;\;\;\;\;\;\;\;\;\;\;\;\;\;\;\;\;\;\;\;\;\;\;\;\;\;\;\;\;\;
\;\;\;\;\;\;\;\;\;\;\;\;\;\;\;\;\;\;\;\;\;\;\;\;\;\;\;\;\;\;\;\;
$$
$$
{\rm He}^0:  
\sigma_{He^0}(\nu)=7.2\times 10^{-18}[1.66(\nu/\nu_{th,{\rm He}^0})^{-2.05}
\;\;\;\;\;\;\;\;\;\;\;\;\;\;\;\;\;\;\;\;\;\;\;\;\;\;\;\;\;\;\;\;\;
$$
$$
\;\;\;\;\;\;\;\;\;\;\;\;\;\;\;\;\;+0.66(\nu/\nu_{th,{\rm He}^0})^{-3.05}]
\;\;\;\;\;\;\;\;\;\;\;\;\;\;\;\;\;\;\;\;\;
$$
$$
{\rm He}^+:  
\sigma_{He^+}(\nu)=1.58\;\times \;10^{-18}(\nu/\nu_{th,{\rm He}^+})^{-3}
\;\;\;\;\;\;\;\;\;\;\;\;\;\;\;\;\;\;\;\;\;\;\;
$$
\\
\item{\it Recombination rates} $\;\;\;$[cm$^{3}$ s$^{-1}$] $\;\;\;\;$(2)\\
$$
{\rm H}^0\;:  \;
\alpha_{H^0}(T)=8.40\times 10^{-11}T^{-1/2}\left(\frac{T}{10^3}\right)^{-0.2}
\left[1+\left(\frac{T}{10^6}\right)^{0.7}\right]^{-1}
$$
$$
{\rm He}^0:  
\alpha_{He^0}(T)=1.50\times 10^{-10}T^{-0.6353}
\;\;\;\;\;\;\;\;\;\;\;\;\;\;\;\;\;\;\;\;\;\;\;\;\;\;\;\;
\;\;\;\;\;\;\;\;\;\;\;\;\;\;\;\;\;\;\;\;\;\;\;\;\;\;\;\;
$$
$$
{\rm He}^+:  
\alpha_{He^+}(T)=3.36\times 10^{-10}T^{-1/2}\left(\frac{T}{10^3}\right)^{-0.2}
\left[1+\left(\frac{T}{10^6}\right)^{0.7}\right]^{-1}
$$
\\
\item{\it Collisional ionization rates} $\;\;\;$[cm$^{3}$ s$^{-1}$] $\;\;\;\;$(2)\\
$$
{\rm H}^0\;:  \;
\gamma_{H^0}(T)=5.85\times 10^{-11}T^{1/2}\left[1+\left(\frac{T}{10^5}\right)^{1/2}\right]^{-1}
e^{-157809.1/T}
$$
$$
{\rm He}^0:  
\gamma_{He^0}(T)=2.38\times 10^{-11}T^{1/2}\left[1+\left(\frac{T}{10^5}\right)^{1/2}\right]^{-1}
e^{-285335.4/T}\
$$
$$
{\rm He}^+:  
\gamma_{He^+}(T)=5.68\times 10^{-12}T^{1/2}\left[1+\left(\frac{T}{10^5}\right)^{1/2}\right]^{-1}
e^{-631515/T}
$$
 The cooling function, $\Lambda(T,n_e,n_{H^0},n_{H^+},n_{He^0},n_{He^+},n_{He^{++}})$, 
is calculated including the contribution of the following radiative processes:\\

\item{\it Collisional ionization cooling} $\;\;\;$[erg cm$^{3}$ s$^{-1}$] $\;\;\;\;$(2)\\
$$
{\rm H}^0 \;: \;
\zeta_{H^0}(T)=1.27\times 10^{-21}T^{1/2}\left[1+\left(\frac{T}{10^5}\right)^{1/2}\right]^{-1}
e^{-157809.1/T}
$$
$$
{\rm He}^0:  
\zeta_{He^0}(T)=9.38\times 10^{-22}T^{1/2}\left[1+\left(\frac{T}{10^5}\right)^{1/2}\right]^{-1}
e^{-285335.4/T}
$$
$$
{\rm He}^+:  
\zeta_{He^+}(T)=4.95\times 10^{-22}T^{1/2}\left[1+\left(\frac{T}{10^5}\right)^{1/2}\right]^{-1}
e^{-631515/T}
$$
\\
\item{\it Recombination cooling} $\;\;\;$[erg cm$^{3}$ s$^{-1}$] $\;\;\;\;$(2)\\
$$
{\rm H}^0\;:  \;
\eta_{H^0}(T)=8.70\times 10^{-27}T^{1/2}\left(\frac{T}{10^3}\right)^{-0.2}
\left[1+\left(\frac{T}{10^6}\right)^{0.7}\right]^{-1}
$$
$$
{\rm He}^0: 
\eta_{He^0}(T)=1.55\times 10^{-26}T^{0.3647}
\;\;\;\;\;\;\;\;\;\;\;\;\;\;\;\;\;\;\;\;\;\;\;\;\;\;\;\;\;
\;\;\;\;\;\;\;\;\;\;\;\;\;\;\;\;\;\;\;\;\;\;\;\;\;\;\;\;\;\;
$$
$$
{\rm He}^+:  
\eta_{He^+}(T)=3.48\times 10^{-26}T^{1/2}\left(\frac{T}{10^3}\right)^{-0.2}
\left[1+\left(\frac{T}{10^6}\right)^{0.7}\right]^{-1}
$$
\\
\item{\it Collisional excitation cooling}  $\;\;\;\;$(2)\\
$$
{\rm H}^0\;:  \;
\psi_{H^0}(T)=7.5\times 10^{-19}\left[1+\left(\frac{T}{10^5}\right)^{1/2}\right]^{-1}
e^{-118348/T}\;\;\;\;
$$
{\small $\;\;\;\;\;\;\;\;[\psi_{H^0}]$=[erg cm$^{3}$ s$^{-1}$] }
$$
{\rm He}^+:  
\psi_{He^+}(T)=5.54\times 10^{-17}T^{-0.397}\left[1+\left(\frac{T}{10^5}\right)^{1/2}\right]^{-1}
e^{-473638/T}
$$
{\small $\;\;\;\;\;\;\;\;[\psi_{He^+}]$=[erg cm$^{3}$ s$^{-1}$] }
$$
{\rm He}^0: 
\psi_{He}(T)=9.10\times 10^{-27}T^{-0.1687}\left[1+\left(\frac{T}{10^5}\right)^{1/2}\right]^{-1}
e^{-13179/T}
$$
{\small $\;\;\;\;\;\;\;\;[\psi_{He}]$=[erg cm$^{6}$ s$^{-1}$] }
\\

{\small The above expressions are for excitations to all H$^0$ levels,	 to $n=2$ for He$^+$ and 
   to the $n=2,3,4$ He$^{0}$ triplets (of the He$^0$($2^3$S) state, supposed to be populated 
by He$^+$ recombinations)}
\\
\item{\it Bremsstrahlung cooling} $\;\;\;$[erg cm$^{-3}$ s$^{-1}$] $\;\;\;\;$(3)\\
$$
\beta(T)=1.42\times10^{-27}T^{1/2}\left[n_{H^+}+n_{He^+}+4n_{He^{++}}\right]n_e
$$
\\
\item{\it Compton cooling/heating} $\;\;\;$[erg cm$^{-3}$ s$^{-1}$] $\;\;\;\;$(4)\\
$$
\varpi(T)=1.017\times10^{-37}T_\gamma^4\left[T-T_\gamma \right]n_e
\;\;\;\;\;\;\;\;\;\;\;\;\;\;\;\;\;\;\;\;\;\;\;\;\;
$$ 
\\ 

References: (1) Osterbrok 1989; (2) Cen 1992; (3) Black 1981; (4) Haiman \etal 1996 
\end{itemize}

\label{lastpage}

\begin{thebibliography}{99}
\bibitem{} Abel, T., Norman, M. L. \& Madau, P. 1999, ApJ, 523, 66  
\bibitem{} Abel, T. \& Wandelt, B. D. 2002, MNRAS, 330, L53
\bibitem{}Benson, A.J., Nusser, A., Sugiyama, N. \& Lacey, C.G. 2001, MNRAS, 320, 153 
\bibitem{}Bianchi S., Ferrara A. \& Giovanardi C., 1996, ApJ, 523,
\bibitem{}Bianchi S., Ferrara A., Davis J. \& Alton P.,2000, MNRAS, 311, 601
\bibitem{}Black, J.H., 1981, MNRAS, 197, 553
\bibitem{}Cashwell, E. D., Evrett, C. J.,1959,{\it A Practical Manual on the Monte Carlo Method 
for Random Walk Problems}, Pergamon, New York
\bibitem{}Cen, R. 1992, ApJS, 78, 341
\bibitem{}Cen, R. 2002, ApJS, 141, 211
\bibitem{}Chiu, W. A. \& Ostriker, J. P. 2000, ApJ, 534, 507
\bibitem{}Ciardi, B., Ferrara, A., Governato, F. \& Jenkins, A. 2000, MNRAS, 314, 611 (CFGJ)
\bibitem{}Ciardi, B., Ferrara, A., Marri, S. \& Raimondo, G. 2001, MNRAS, 324, 381 (CFMR)
\bibitem{}Ciardi, B., Stoehr, F. \& White, S.D.M. 2003, astro-ph/0301293 (CSW)
\bibitem{}Fan, X. et al. 2001, AJ, 122, 2833
\bibitem{}Fan, X. et al. 2003, astro-ph/0301135
\bibitem{}Ferrara, A., Bianchi, S., Dettmar, R. J. \& Giovanardi, C., 1996, ApJ, 467, L69
\bibitem{}Ferrara, A., Bianchi, S., Cimatti, A. \& Giovanardi, C., 1999, ApJS, 123, 437
\bibitem{}Gnedin, N. Y. 2000, ApJ, 535, 530
\bibitem{}Gnedin, N. Y. \& Abel, T. 2001, NewA, 6, 437
\bibitem{}Gnedin, N. Y. \& Ostriker, J. P. 1997, ApJ, 486, 581
\bibitem{}Haiman, Z., Thoul, A. A., Loeb, A. 1996, ApJ, 464, 523
\bibitem{}Haiman, Z. \& Loeb, A. 1997, ApJ, 483, 21
\bibitem{}Hu, E.M. et al. 2002, ApJ, 576, 99
\bibitem{}Mihalas, D., 1978, {\it Stella Atmospheres}, W. H. Freeman and Company, San Francisco
\bibitem{}Miralda-Escud\'e, J., Haehnelt, M. \& Rees, M. R. 2000, ApJ, 530, 1
\bibitem{}Osterbrok, D. E., 1989, {\it Astrophysics of Gaseous Nebulae and Active Galactic Nuclei}, Mill Valley: University science Books 
\bibitem{}Razoumov, A. O., Norman, M. L., Abel, T. \& Scott, D. 2002, ApJ, 572, 695
\bibitem{}Razoumov, A. \& Scott, D. 1999, MNRAS, 309, 287
\bibitem{}Sokasian, A., Abel, T. \& Hernquist, L. E. 2001, NewA, 6, 359
\bibitem{}Spitzer, L. Jr., 1978, {\it Phisycal Processes in the Interstellar Medium}, Jonh Wiley \& Sons, New York
\bibitem{}Umemura, M., Nakamoto, T. \& Susa, H. 1999, in Numerical Astrophysics, 
 eds.Miyama et al. (Kluwer: Dordrecht), p.43

\bibitem{}Valageas, P. \& Silk, J. 1999, A\&A, 347, 1
\end{thebibliography}
\end{document}